\documentclass[fleqn,usenatbib]{mnras}
\usepackage[T1]{fontenc}

\DeclareRobustCommand{\VAN}[3]{#2}
\let\VANthebibliography\thebibliography
\def\thebibliography{\DeclareRobustCommand{\VAN}[3]{##3}\VANthebibliography}
\def\nodata{$\dots$}

\usepackage{graphicx}	% Including figure files
\usepackage{amsmath}	% Advanced maths commands
\usepackage{amssymb}	% Extra maths symbols
\usepackage{pdflscape}	% Landscape pages

\usepackage{epsfig}
\usepackage{epstopdf}
\usepackage{multirow}
\usepackage{hyperref}
\usepackage{upgreek}
\title[LWA Pulsar Polarimetry]{Detecting Pulsar Polarization below 100 MHz with the Long Wavelength Array}

\author[V. Dike et al.]{V. Dike,$^{1,2}$,
G. B. Taylor,$^{1}$
J. Dowell,$^{1}$, and
K. Stovall,$^{1,3}$
\\
$^{1}$Department of Physics and Astronomy, University of New Mexico, Albuquerque, NM 87131, USA\\
$^{2}$Department of Physics and Astronomy, University of California Los Angeles, Los Angeles, CA, 90095, USA\\
$^{3}$Sandia National Laboratories, Albuquerque, NM, 87185, USA\\
}

\date{Accepted 2020-Jun-13. Received 2020-May-22; in original form 2020-Apr-01}

\pubyear{2020}

\begin{document}
\label{firstpage}
\pagerange{\pageref{firstpage}--\pageref{lastpage}}
\maketitle

\begin{abstract}
	
Using the first station of the Long Wavelength Array (LWA1), we examine polarized pulsar emission between 25 and 88 MHz. Polarized light from pulsars undergoes Faraday rotation as it passes through the magnetized interstellar medium. Observations from low frequency telescopes are ideal for obtaining precise rotation measures (RMs) because the effect of Faraday rotation is proportional to the square of the observing wavelength. With these RMs, we obtained polarized pulse profiles to see how polarization changes in the 25--88 MHz range.  The RMs were also used to derive values for the electron density weighted average Galactic magnetic field along the line of sight. We present rotation measures and polarization profiles of 15 pulsars acquired using data from LWA1. These results provide new insight into low-frequency polarization characteristics and pulsar emission heights, and complement measurements at higher frequencies. 
\end{abstract}

\begin{keywords}
pulsars: general -- techniques: polarimetric -- radio telescopes -- ISM: magnetic fields -- Galaxy: structure
\end{keywords}

\section{Introduction}\label{sec:intro}

Shortly after the discovery of the first pulsar, \cite{LyneSmith1968} established that some pulsars are substantially linearly polarized and suggested using a measure of Faraday rotation of the polarized signal to study the magnetic field of the intervening interstellar medium. Faraday rotation is the rotation in the polarization position angle of the source's light as it travels through the magnetized interstellar medium. This rotation occurs because the left- and right-circularly polarized components of the light have different indices of refraction. The proportionality constant is the rotation measure (RM), an observable that can be used to derive the density weighted average component of the magnetic field parallel to the line of sight from the detector to the pulsar, $ B_\|$:
\begin{equation}	
{\rm RM} = \dfrac{e^3}{2\pi m_{\rm e}c^4} \int_{0}^{L}n_e B_\| ds ,
\end{equation}
where $e$ is the electron charge, $m_e$ is the electron mass, $c$ is the speed of light in vacuum, and $n_{\rm e}$ is the electron density. The pulsar's dispersion measure (DM) can be expressed as ${\rm DM}= \int_{0}^{L}n_{\rm e} ds$. Thus, the weighted average magnetic field parallel to the line of sight is proportional to ratio to the RM value to the DM value.

This simple relation between magnetic field strength and RM and DM make pulsars indispensable for Galactic magnetic field measurements \citep{Han2013}. Pulsar RMs were first exploited to measure Galactic magnetic fields by \cite{Manchester1972} at 410 MHz. More recent measurements have been made by \cite{RMFIT2008} and \cite{Han2018} at 1.5 GHz with the Parkes Observatory, and by \cite{sobey2019} at 150 MHz using LOFAR. Faraday rotation is proportional to ${\lambda^2}$, so at lower observing frequencies where the effect of Faraday rotation is greater, RMs can be obtained with greater precision. The limiting factor for both LOFAR and LWA observations is the ability to correct for the Faraday rotation induced by the Earth's ionosphere, typically between 1 and 3 rad m$^{-2}$ \citep{mal18}.

Another reason to measure RMs in the long wavelength regime is to study the properties of pulsar emission. Polarization profiles show a change in morphology of beam structure with frequency. Linear and circular polarization tend to decrease at shorter wavelengths. One theory for this wavelength-dependent polarization property was put forth by \cite{Depolar1998}, assuming a magnetosphere of cold, low-density plasma: the pulsar beam has orthogonal propagation modes that superpose due to birefringence and thus lose polarization at higher frequencies. This theory can be tested by comparing high and low frequency observations to see how linear and circular polarization changes (as in \citealt{LOFAR2015}).

Pulsar geometry can be modeled based on polarimetry. The observed change in polarization position angle (PA) can be related to the physical geometry of the pulsar --- the angles between the spin axis, the beam, and the line of sight --- using the model put forth in \cite{RadhaCooke1969}, typically referred to as the Rotating Vector Model (RVM). For more on RVM fitting as well as a comparison of different beam models, see \cite{EW2001}. Using RVM fitting and the PA data, pulsar emission heights can be estimated. \cite{BCW1991} added special relativistic effects to the RVM to show that the emission height can be approximated as $h \approx  {\rm c} \Delta t/4$, where $\Delta t$ is the delay between the center of polarization position angle curve and the center of the pulse profile. In the interpretation of radius-to-frequency mapping \citep{RandS1975}, different observing frequencies correspond to different emission heights above the pulsar.

Observations of pulsar rotation measures at low frequencies can also be used to explore the density structure of the Earth's ionosphere \citep{mal18}.  Assuming that the RM of the pulsar is known, along with a model for the Earth's magnetic field, one can {interpret} the fluctuations of the RM over time to changes in the density of the ionosphere with time. 

The multitude of reasons to examine low-frequency pulsar polarization have motivated the recent studies of \cite{Johnston2008} at  243,  322  and 607 MHz with the Giant Metrewave Radio Telescope (GMRT), \cite{ankin2010} at frequencies centered between 49-1400 MHz with Arecibo Observatory, \cite{LOFAR2015} and \cite{sobey2019} at 150 MHz with the Low-Frequency Array (LOFAR), and \cite{Mitra2016} at 325 and 610 MHz with the GMRT. Using LWA1, we are able to complement these studies at observing frequencies between 25 and 88 MHz. We organize our results into four categories: our low-frequency RMs for fifteen pulsars; using these RMs, our derivations of the average Galactic magnetic field along the line of sight; the polarized pulse profiles for each pulsar; and our estimates for the height of emission at each frequency tuning.

\section{Observations}\label{sec:observations}
We observed with the first station of the Long Wavelength Array \citep[LWA1;][]{LWA2012}, a 256-antenna dipole array co-located with the Karl G. Jansky Very Large Array on the Plains of San Agustin in New Mexico. The dipoles are arranged in a pseudo-random configuration within an ellipse that is 110 m by 100 m and elongated in the north-south direction. LWA1 is able to observe with four independent beams simultaneously. Observations were performed between July 2015 and August 2016. Pulsars were selected for this study if they are particularly bright in the LWA frequency band or if there was evidence of polarization from a plot of Stokes' parameters. {No polarization calibration was applied to the data as the instrumental polarization for the LWA1 station is quite small at angles within 45 degrees of the zenith. \cite{obe15} find the instrumental polarization to be less than 10\% for linear polarization and less than 5\% for circular polarization.}  Fifteen pulsars were observed with LWA1's beamforming mode for an hour each. Two beams were used to observe each source; each beam is split into two spectral tunings centered on different frequencies. This produced observations in four frequency ranges, each with a bandwidth of 19.6 MHz. The four tunings were centered at 35.1, 49.8, 64.5, and 79.8 MHz. The raw data was coherently de-dispersed and folded using the program DSPSR\footnote{\url{http://dspsr.sourceforge.net/}} \citep{DSPSR2011}. Pulsars are routinely observed with LWA1 and de-dispersed data files, including the observations analyzed in this paper, are available to the public at the LWA Pulsar Data Archive\footnote{\url{http://lda10g.alliance.unm.edu/PulsarArchive/}}. For more information about LWA pulsar data, see \cite{sto15}. 

%% POSSIBLY COMBINE OBSERVATIONS AND METHODS INTO ONE SECTION

\section{Methods}\label{sec:methods}

 Data reduction made use of PSRCHIVE (\citealt{PSRCHIVE2004}, \citealt{PSRCHIVE2012}), a pulsar data analysis software package. Radio frequency interference (RFI) was removed using a median filtering algorithm and manually through visual inspection. The rotation measures were calculated using the method developed by \cite{RMFIT2008}.  The change in total linear polarization was calculated for 512 to 1024 trial RMs near the value obtained at higher observing frequency previously published in the ATNF Pulsar Database\footnote{\url{http://www.atnf.csiro.au/people/pulsar/psrcat/}} \citep{ATNF2005}. The RM for each observation was found using the program RMFIT which fits a Gaussian to a plot of the peak linearly polarized emission as a function of the trial RM. {Uncertainties in the RM were determined using a 68\% confidence interval.}  We corrected the linear polarization of the pulsar for the effect of {Faraday rotation} in order to produce polarization profiles. The effect of Faraday rotation through the ionosphere had to be subtracted to derive the Galactic contribution to the RM.  This was done using the model developed by \cite{Soto2013}. Faraday rotation through the heliosphere was not accounted for, but this contribution is expected to be small because our observations were made at large angles (greater than 99$\degr$) from the Sun, with the exceptions of PSR B0943+10 and PSR B1929+10, for which significantly polarized observations were not available at larger angles (see \ref{subsec:ppp} for details).
    
In order to find the emission heights using the RM-corrected polarization data, we use the RVM calculated with the PSRCHIVE tool {\tt psrmodel}\footnote{\url{http://psrchive.sourceforge.net/manuals/psrmodel/}}, which fits the PA curve to the RVM:
    
    \begin{equation}	
    \tan(\psi - \psi_0)= \dfrac{\sin \alpha \sin(\phi -\phi_0)}{\sin \zeta \cos \alpha - \cos \zeta \sin \alpha \cos (\phi - \phi_0)} , 
    \end{equation}
\noindent 
\phantom{break}\\
where $\psi$ is the PA, $\alpha$ is the angle between the spin and magnetic axes, $\zeta$ is the angle between the spin axis and the line of sight, and $\phi$ is the phase of the pulsar. The angles $\psi_0$ and $\phi_0$ refer to the PA at the point of highest rate of change and the corresponding value of $\phi$ at that point, respectively. To get a value for the pulsar phase in the middle of the Stokes' $I$ profile corresponding to the emission height, $\phi_{\rm em}$, we used one of three assumptions:
\begin{enumerate}
	\item  $\phi_{\rm em}$ is the midpoint between the start and end of the {on-pulse window} for pulsars that had a single, broad peak.
	\item $\phi_{\rm em}$ is the phase corresponding to the highest point of the peak for pulsars with a single, narrow peak.
	\item $\phi_{\rm em}$ is the phase corresponding to lowest point of the central valley for double-peaked pulsars.
\end{enumerate}
The phase difference, $\Delta \phi$, is then defined as $\phi_0 - \phi_{\rm em}$. We put this value into the equation for emission height from \cite{BCW1991}:

\begin{equation}
    h \approx \frac{1}{4} \frac{{\rm{c}}P \Delta \phi }{2 \pi},
\end{equation}
where $P$ is the period of the pulsar.  This is effectively the light cylinder radius multiplied by $\Delta \phi$.

As an alternative the emission height can be constrained using the pulse width and geometric arguments assuming a dipolar field.  This leads to a height, $h_{\rm geo}$, of:

\begin{equation}	
    h_{\rm geo} = \frac{4}{9}\frac{{\rm c}P}{2\pi}\left[\frac{w^2_{10}}{4}\sin{\alpha}\sin{(\alpha+\sigma)} + \sigma^2\right], 
    \end{equation}
\noindent 
\phantom{break}\\
where $w_{10}$ is the pulse width at 10\% of the maximum and $\sigma = \zeta - \alpha$ \citep{Phillips92}.

\section{Results}\label{sec:results}

\subsection{Rotation Measures}\label{subsec:rm}
We were successful in obtaining a rotation measure in at least one frequency tuning for 15 pulsars. Table \ref{bigtable} lists the results as well as the period and DM of each pulsar. The few cases where RMs were unable to be found at certain frequencies are usually because of lower sensitivity in the lowest frequency part of the band, depolarization at lower frequencies, or RFI in that band. We note that although the uncorrected RM values have errors typically less than 0.03 rad m$^{-2}$, the ionospheric corrections add considerable uncertainty to the final RM values. 

 \begin{table*}
  \caption{Results of the RM fitting process for 15 pulsars. The period is taken from the ATNF Pulsar Database and the DM is from \citep{sto15}. The frequency column lists the center frequency of the beam used for each observation. Column 5 contains the RM measurement taken directly from the observation while column 6 is the ionospheric RM at the time of observation. The rightmost column is the observed RM with the ionospheric RM subtracted.  Uncertainties in the last decimal place are indicated in parenthesis.}
  \label{bigtable}
  \begin{tabular}{cccccccccc}
    \hline
    Pulsar Name & {Epoch} & Period & DM & Frequency & RM & Ionosphere RM & Corrected RM\\
     & {(MM-DD-YYYY)} & (s) & (pc cm$^{-3}$) & (MHz) & (rad m$^{-2}$) & (rad m$^{-2}$) & (rad m$^{-2}$) \\
    \hline
	{B0329+54} & 01-26-2016 & {0.7145} & {26.779(1)} & 64.5 & $-$63.776(4) & {2.0(2)} & $-$65.8(2)  \\  
	& & &	& 79.2 & $-$63.780(4) & & $-$65.8(2) \\ 
		\hline		   
	   B0628$-$28  & 12-24-2015 & {1.2444} & {34.425(1)} & 49.8 & 46.67(5) & {2.3(4)} & 44.4(4) \\
	& & &	& 64.5 & 46.676(6) & & 44.4(4) \\
	& & &	& 79.2 & 46.69(1)  & & 44.4(4) \\ 		
	\hline		
	{B0809+74} & 01-14-2016 & {1.2922} & {5.771(2)}  & 49.8 & $-$13.33(1) &  {1.5(2)} & $-$14.8(2) \\
	& & &	& 64.5 & $-$13.35(3) & & $-$14.8(2) \\
	& & &	& 79.2 & $-$13.36(3) & & $-$14.8(2) \\ 
	\hline						   
	{B0823+26} & 01-26-2016 
	& {0.5307} & {19.4789(2)}	& 79.2 & 5.95(2)  &  {1.3(2)} & 4.6(2)\\ \hline	
	{B0834+06} & 01-27-2016 & {1.2738} & {12.8640(4)}  & 49.8 & 26.15(1) & {1.6(2)} & 24.6(2) \\
	& & &	& 64.5 & 26.169(6) & & 24.6(2) \\
	& & &	& 79.2 & 26.129(4) & & 24.6(2) \\ 
	\hline						   
	{B0919+06} & 03-26-2016 & {0.4306}  & {27.2986(5)} & 49.8 & 33.96(3) & {1.3(2)} &32.7(2) \\
	&   & &	& 64.5 & 33.960(4) & &32.7(2) \\
	&   & &	& 79.2 & 33.90(2) & &32.6(2) \\ 
	\hline
	{B0943+10$^a$} & 08-10-2016 
	& {1.0977} & {15.334(1)}	& 64.5 & 16.07(2)   & {3.9(2)} & 12.2(2) \\
	&   & &	& 79.2 & 16.3(2) & & 12.4(3)\\ 
	\hline  
	{B0950+08}  & 01-26-2016  & {0.2531} & {2.96927(8)}	& 35.1 & 2.210(2) & {1.5(2)} & 0.7(2) \\
	& & &	& 49.8 & 2.2111(4) & &0.7(2) \\
	& & &	& 64.5 & 2.209(1) & &0.7(2)\\
	& & &	& 79.2 & 2.215(3) & &0.8(2)\\ \hline						   	   
	{B1133+16}  & 01-14-2016 & {1.1879} & {4.8480(2)} 	& 35.1 & 4.71(4) & {1.2(2)} & 3.5(2) \\ 
	&  & & & 49.8 & 4.710(1) & & 3.5(2)\\
	&  & &	& 64.5 & 4.713(3) & & 3.5(2)\\
	&  & &	& 79.2 & 4.719(6) & & 3.5(2)\\ 	
	\hline
	{B1604$-$00} & 03-26-2016 & {0.4218}   & {10.6823(1)}	& 35.1 & 7.13(4) & {1.4(3)} & 5.7(3) \\
	&   & &	& 49.8 & 7.115(3) &  & 5.7(3)\\
	&   & &	& 64.5 & 7.10(1) &  & 5.7(3)\\ \hline					   
	{B1822$-$09} & 07-10-2016 & { 0.7690 } & {19.3833(9)}	& 79.2 & 68.9(3)  & {1.8(2)} & 67.1(3) \\ 
	\hline
	{B1839+56} & 06-18-2016 & {1.6529}  & {26.774(1)}  & 49.8 & $-$3.25(2) & {1.3(2)} & $-$4.5(2) \\	
	%& 64.5 & -3(4) & \\
	&   & &	& 79.2 & $-$3.3(1) & & $-$4.6(2) \\ 
	\hline
	{B1919+21} & 07-15-2015 & {1.3373}  & {12.4386(3)}	& 64.5 & $-$16.078(5) & {1.4(2)} &$-$17.4(2) \\
	&  & &	& 79.2 & $-$16.091(8) & &$-$17.4(2) \\ 
	\hline						   
	{B1929+10$^a$} & 02-10-2016
	&{0.2265} & {3.1828(5)}	& 79.2 & $-$6.5(5)$^b$    & {3.4(2)} & $-$9.9(5) \\ 
	\hline
	{B2217+47} & 08-08-2015 & {0.5385}  & {43.4975(5)} 	& 64.5 & $-$35.07(1) & {1.2(2)} & $-$36.3(2) \\ 
	& & & & 79.2 & $-$35.063(6) & & $-$36.3(2) \\
    \hline
  \end{tabular}\\
  $^a$This pulsar was observed close to the influence of the heliosphere.\\
$^b$ This error was estimated from visually inspecting the graph of trial rotation measure vs. linear polarization because Gaussian curve fitting failed.
 \end{table*}

In general the results reported in Table \ref{bigtable} are in excellent agreement with the 8 (mostly northern) pulsars mutually observed by us and \cite{sobey2019}.  The mean difference between this set of 8 low-frequency pulsar RMs is $-$1.02 $\pm$ 0.83.  The range of differences runs from $-$0.35 for B1919$+$21 to $-$2.76 rad m$^{-2}$ for B1929$+$10.  The average difference of $-$1.02 rad m$^{-2}$ appears significant, and is most likely due to differences in correcting for the ionosphere.  The difference is unlikely to arise from the modest difference in frequency between the \cite{sobey2019} and our own, since our observations of the RM which span the frequency range from 25 to 88 MHz reveal no systematic change in RM as a function of frequency to less than 0.1 rad m$^{-2}$. {Although \cite{sobey2019} used an RM synthesis technique and we used a technique that maximizes linearly polarized emission, neither technique is expected to have a systematic bias.}

\subsection{Average Magnetic Field}\label{subsec:bfield}

The electron density weighted average magnetic field parallel to the line of sight can be expressed using RM and DM as defined in section \ref{sec:intro}:

\begin{equation}
\langle B_\| \rangle=\dfrac{\int_{0}^{L}n_e B_\| ds}{\int_{0}^{L}n_e ds}=\dfrac{2\pi m_{\rm e}c^4}{e^3}\dfrac{\rm RM}{\rm DM},
\end{equation}
where the constants can be evaluated and expressed in the convenient units of cm$^{3}$ m$^{-2}$ pc$^{-1}$ $\upmu$G$^{-1}$ to match the standard units of RM and DM, resulting in a magnetic field in $\upmu$G . 

\begin{figure*}
	\includegraphics{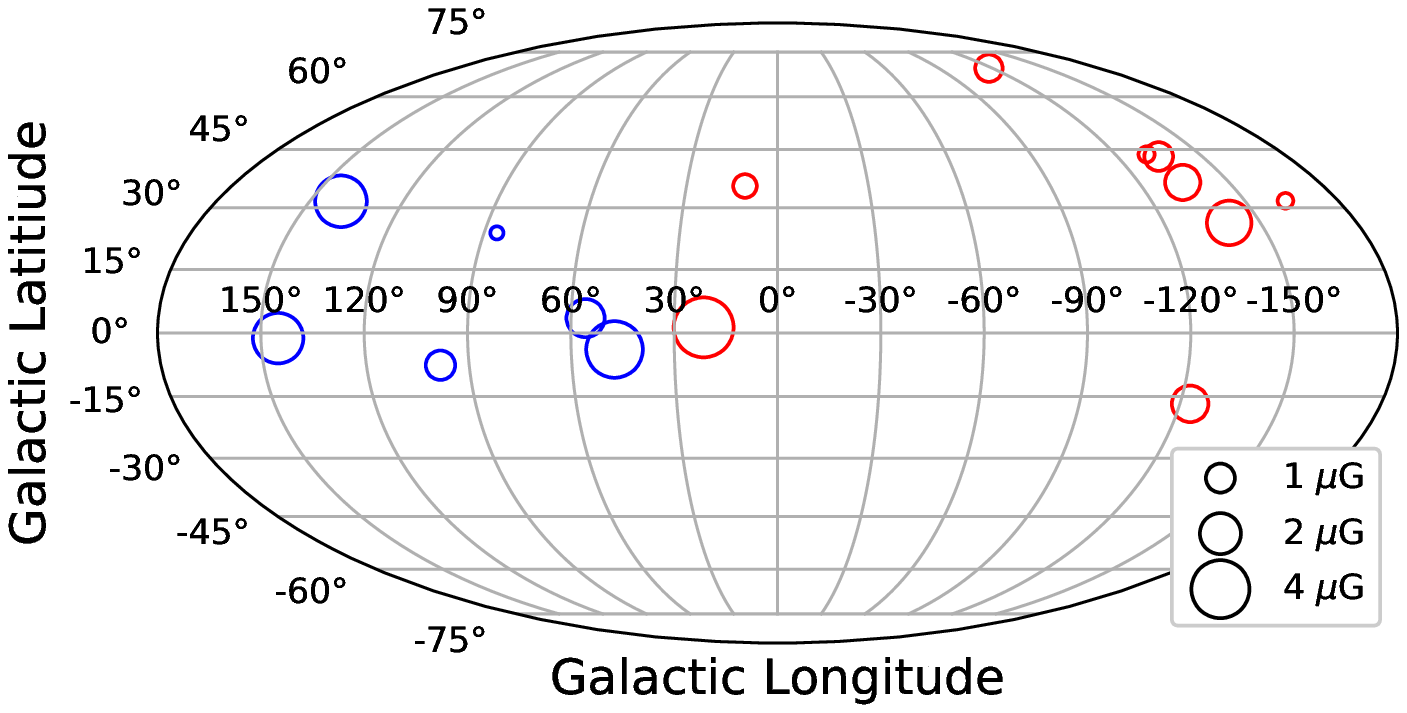}
	\caption{Derived Galactic magnetic field. Red circles correspond to a positive $\langle B_\| \rangle$, while blue corresponds to negative $\langle B_\| \rangle$. The size of the circles is proportional to the absolute value of the magnetic field strength, ranging from 0.21 $\upmu$G to 4.26 $\upmu$G.\label{bfieldmap}}
\end{figure*}

Table \ref{smalltable} shows the values we derived for $\langle B_\| \rangle$ from our RMs. This data is plotted in Galactic coordinates in Figure \ref{bfieldmap}. In the map, there is a trend of positive magnetic field (and rotation measure) for Galactic longitudes less than $30\degr$ and negative magnetic field for Galactic longitude greater than $30\degr$. The higher dispersion at lower frequencies limits our pulsar sample to those relatively nearby; all the pulsars in this study are within 3 kpc, and all but three are within 1 kpc, and show the structure of the magnetic field between 0.2 and 1 kpc from the Sun. This region is thought to be in between the denser spiral arms of the Galaxy and has a clockwise magnetic field \citep{Han2006}. Our map is consistent with the work of \cite{Manchester1972} and \cite{sobey2019}. It should be noted that clouds of ionized hydrogen in between the pulsar and Earth can confuse magnetic field measurements from pulsars \citep{Mitra2003}, but these clouds are fairly rare and are not expected to affect our measurements.
\cite{frisch2012} explore the region within 40 parsecs of the Sun using polarized light from stars.  They suggest a scenario where the local magnetic field is connected to the Loop I superbubble.  This nearby structure could explain the dichotomy in magnetic field distribution that we see in Fig.~\ref{bfieldmap} near $l\sim$30$^\circ$.  We also note that this area where the sign of $\langle B_\| \rangle$ changes appears to roughly correspond to a depolarization area identified by \citet{Wolleben07} at 1.4 GHz, suggesting that this feature is within a few hundred pc.

 \begin{table}
 \centering
  \caption{Values of $\langle B_\| \rangle$ derived from the values of RM and DM in table \ref{bigtable}. The values for distance, Galactic longitude ($l$), and Galactic latitude ($b$) are from the ATNF pulsar database.}
  \label{smalltable}
  \begin{tabular}{ccccc}
\hline
{Pulsar Name} & {Distance} & $l$ & $b$ & {$\langle B_\| \rangle$} \\
		&  (kpc) & {($\degr$)}  & ($\degr$) & ($\upmu$G) \\
\hline
	{B0329+54} & 1.00 & 145.00 & -1.22 & -3.024(9)   \\    %	
	{B0628-28} & 0.32 & -123.05 & -16.76 & 1.59(1) \\ 		 %	
	{B0809+74} & 0.43 & 140.00 & 31.62 & -3.17(5)  \\  %						   
	{B0823+26} & 0.32 & -163.04 & 31.74 & 0.29(1)   \\  %	
	{B0834+06} & 0.19 & -140.28 & 26.27 & 2.35(2)   \\  %						   
	{B0919+06} & 1.10 & -134.58 & 36.39 & 1.474(8)   \\  %
	{B0943+10} & 0.89 & -134.59 & 43.13 & 0.99(3)     \\  %  
	{B0950+08} & 0.26 & -131.09 & 43.70 & 0.31(7)   \\  %   
	{B1133+16} & 0.35 & -118.10 & 69.20 & 0.89(6)   \\ 	 %		   
	{B1604$-$00} & 0.68 & 10.72 & 35.47 & 0.66(3)    \\  %
	{B1822$-$09} & 0.30 & 21.45 & 1.32 & 4.26(2)   \\  %
	{B1839+56} & 2.19 & 86.08 & 23.82 & $-$0.21(1)    \\  %
	{B1919+21} & 0.30 & 55.78 & 3.50 & -1.73(2)   \\  %						   
	{B1929+10} & 0.31 & 47.38 & -3.88 & -3.8(2)   \\  %	 
	{B2217+47} & 2.39 & 98.38 & -7.60 & -1.029(5)   \\ 
	\hline
	\end{tabular}
\end{table}

\subsection{Polarized Pulse Profiles}\label{subsec:ppp}

Using the RM to correct the linear polarization, we found a total of 37 pulse profiles. These profiles are 
plotted in Figures 2-16.  Each plot shows polarization in the bottom section, with black corresponding to total intensity, while red is  linearly polarized intensity, and blue is {circularly polarized intensity}: Stokes' $I$, $P$ ($\sqrt{{Q}^2+{U}^2}$), and $V$, respectively. The change in PA in degrees vs. pulse phase is plotted in the top portion. For small fractional polarization ($P/I<0.3$), the position angle is not plotted.

There are many interesting features that can be discerned from our low-frequency polarized profiles. As noted in Section \ref{sec:intro}, some pulsar emission models predict depolarization at low frequencies, as can be seen clearly in the case of PSR B0628-28 (Figure \ref{B0628fig}) and PSR B0919+06 (Figure \ref{B0919fig}) especially when compared the profiles created by \cite{1998GouldLyne} at 410 MHz and by \cite{Johnston2008} at 243 MHz.

For PSR B0950+08 (Figure \ref{B0950fig}) we find an unusually high fraction of linear polarization in our frequency band compared to other pulsars studied here, approaching 100\% at 64.5 and 79.2 MHz.  There is some depolarization at 49.8 MHz. This pulsar is known to have an ``interpulse," a small pulse far from the main pulse, at higher frequencies \citep{interpulse1968}, but we do not see the interpulse in our profiles. PSRs B0823+26 (Figure \ref{B0823fig}), B0834+06 (Figure \ref{B0834fig}), and B1133+16 (Figure \ref{B1133fig}) show an unusually high degree of circular polarization. 

The only significantly polarized observation of B1929+10 was unfortunately taken only 40$\degr$ from the Sun, resulting in a noisy profile. Similarly, the unusual appearance of B0943+10 (Figure \ref{B0943fig}) is because of our observing it 8$\degr$ from the Sun, during the pulsar's burst-mode. 

PSR B1822$-$09 (Figure \ref{B1822fig}), another mode-switching pulsar, has a two-component pulse profile in its {B-mode} at higher frequencies, which can be seen in the profiles of \cite{1998GouldLyne}, but at lower frequencies the leading component diminishes \citep{Johnston2008}. The only profile we obtained for this pulsar, at 79.2 MHz, does not have the leading component.

Several profiles have double peaks that change separation depending on frequency. The peaks are generally observed to get farther apart at lower frequencies (higher emission heights). The usual widening can be seen in PSRs B0809+74 (Figure \ref{B0809fig}), B0943+10 (Figure \ref{B0943fig}), B0950+08 (Figure \ref{B0950fig}), and B1133+16 (Figure \ref{B1133fig}). However, PSR B1919+21 (Figure \ref{B1919fig}) does not obey this rule. 

We also found double-peaked structure in pulsars that have a single peak at higher frequencies. The double-peaked structure of B0809+74 is apparent below 79.2 MHz, but seems to merge at the observation of \cite{LOFAR2015} at 150.9 MHz. The slight splitting seen in observations above 400 MHz in the profile of PSR B1604$-$00 (Figure \ref{B1604fig}) becomes a clear separation in our profiles as well as in \cite{ankin2010}. but at 35.1 MHz the peaks seem to come together. Note that at our frequencies, PSR B0919+06 does not show a continuation of the bifurcation that \cite{Johnston2008} discussed for 322 and 243 MHz.

\section{Emission Heights}\label{sec:eh}

We calculated emission heights for the eight pulsars in our sample with peak polarization fractions greater than 30\%.  Table \ref{othertable} shows our results for $\alpha$, $\zeta$, $\Delta \phi$, $w_{10}$, and the emission heights as a function of frequency.  Values of $\alpha$ or $\zeta$ that are poorly constrained by the fits, i.e., when the uncertainty exceeds 90$^\circ$, have been omitted.  Two of the pulsars in our sample, B1822$-$09 and B2217$+$47, also had poorly constrained values of $\Delta \phi$ in addition to $\alpha$ and $\zeta$ and no emission height could be determined.  In general we find that the values of $\alpha$ and $\zeta$ are closely aligned, indicating that the magnetic axis lies near the line of sight, and the opening angle of the cone of emission is small.

 \begin{table*}
 \centering
  \caption{Results for the Radius-Frequency Mapping.\label{othertable}}
  \begin{tabular}{cccccccc}
\hline
{Pulsar Name} & {Frequency} & $\alpha$ & $\zeta$ & {$\Delta \phi$} & $w_{10}$ & $h$ & $h_{\rm geo}$ \\
 & (MHz) & ($^\circ$) & ($^\circ$) & ($^\circ$) & ($^\circ$) & (km) & (km) \\
 \hline
 {B0329$+$54} & 64.5 & \nodata & \nodata & 5.8 $\pm$ 4.0 & 24.8 $\pm$ 0.7 & 860 $\pm$ 600 & \nodata \\
{B0329$+$54} & 79.2 & \nodata & \nodata & 0.8 $\pm$ 2.5 & 9.1 $\pm$ 0.7 & 120 $\pm$ 370 & \nodata \\
{B0628$-$28} & 49.8 & 130.1 $\pm$ 77.6 & 139.7 $\pm$ 66.3 & 3.1 $\pm$ 6.5 & 37.3 $\pm$ 1.4 & 810 $\pm$ 1680 & 4780 $\pm$ 100 \\
{B0628$-$28} & 64.5 & \nodata & \nodata & 1.0 $\pm$ 3.9 & 35.9 $\pm$ 1.4 & 250 $\pm$ 1000 & \nodata \\
{B0628$-$28} & 79.2 & 64.3 $\pm$ 9.8 & 76.1 $\pm$ 10.9 & 1.5 $\pm$ 3.0 & 34.1 $\pm$ 1.4 & 400 $\pm$ 790 & 7080 $\pm$ 170 \\
{B0919$+$06} & 49.8 & \nodata & \nodata & 9.2 $\pm$ 1.4 & 31.3 $\pm$ 0.7 & 830 $\pm$ 120 & \nodata \\
{B0919$+$06} & 64.5 & \nodata & \nodata & -4.7 $\pm$ 4.7 & 28.1 $\pm$ 1.4 & -420 $\pm$ 420 & \nodata \\
{B0950$+$08} & 35.1 & \nodata & \nodata & -5.2 $\pm$ 7.9 & 49.2 $\pm$ 8.4 & -270 $\pm$ 420 & \nodata \\
{B0950$+$08} & 49.8 & 138.2 $\pm$ 37.9 & 157.7 $\pm$ 23.1 & -2.2 $\pm$ 2.2 & 61.9 $\pm$ 2.8 & -120 $\pm$ 110 & 2280 $\pm$ 40 \\
{B0950$+$08} & 64.5 & \nodata & 161.7 $\pm$ 57.2 & -4.3 $\pm$ 2.8 & 57.0 $\pm$ 1.4 & -220 $\pm$ 150 & \nodata \\
{B0950$+$08} & 79.2 & 76.2 $\pm$ 33.9 & 104.7 $\pm$ 39.4 & -4.2 $\pm$ 2.2 & 49.9 $\pm$ 1.4 & -220 $\pm$ 120 & 5140 $\pm$ 50 \\
{B1133$+$16} & 49.8 & \nodata & \nodata & 3.1 $\pm$ 0.4 & 18.3 $\pm$ 0.7 & 780 $\pm$ 90 & \nodata \\
{B1133$+$16} & 64.5 & \nodata & \nodata & 1.5 $\pm$ 0.7 & 16.5 $\pm$ 0.7 & 380 $\pm$ 160 & \nodata \\
{B1133$+$16} & 79.2 & \nodata & \nodata & 2.7 $\pm$ 0.4 & 15.8 $\pm$ 0.7 & 660 $\pm$ 110 & \nodata \\
{B1604$-$00} & 35.1 & 48.2 $\pm$ 10.4 & 114.1 $\pm$ 19.6 & -4.9 $\pm$ 7.7 & 15.5 $\pm$ 2.8 & -430 $\pm$ 680 & 26810 $\pm$ 40 \\
{B1604$-$00} & 49.8 & 69.0 $\pm$ 6.6 & 81.8 $\pm$ 7.3 & 3.1 $\pm$ 1.5 & 12.7 $\pm$ 2.8 & 270 $\pm$ 130 & 1240 $\pm$ 40 \\
{B1604$-$00} & 64.5 & 79.5 $\pm$ 2.4 & 89.2 $\pm$ 2.9 & 3.8 $\pm$ 1.9 & 11.2 $\pm$ 2.8 & 340 $\pm$ 170 & 770 $\pm$ 40 \\
\hline
	\end{tabular}
\end{table*}

We also find the heights calculated through the RVM method to be lower than those from the geometric model with several of the RVM values being consistent with zero height.  The average emission height for our sample is 130 $\pm$ 90 km at 64.5 MHz.  For three of our pulsars (B0919+06, B0950+08, and B1133+16) that were analyzed by \cite{1997Effels} in a similar fashion at 1400 MHz they find an average emission height of 310 $\pm$ 230 km. Similarly, \cite{BCW1991} derived average heights for 18 pulsars at 1418 MHz of 300 $\pm$ 200 km.  These results are in contrast to the expected trend of an inverse relationship between emission height and frequency.  Indeed, \cite{1997Effels} fit a power law to 11 pulsars observed between 430 and 1400 MHz and found $a$=$-$0.3 $\pm$ 0.1.  Similarly, studies by \cite{sto15} and \cite{pil16} of component widths and spacings generally find increasing sizes and separations with lower frequencies, with some exceptions.

This apparent discrepancy could be the result of poorly constrained fits to the RVM model.  Examining Figures 2-16 we see that most of the polarized emission occurs in a relatively small window of pulsar phase.  This, combined with the lower signal-to-noise ratio of the polarized intensity and the complex functional form of the RVM model, leads to larger uncertainties in the fitted parameters.  This is clearly the case for $\alpha$ and $\zeta$ but the values of $\phi_0$ are better constrained as can be seen by the uncertainties on $\Delta \phi$.  Although there is insufficient overlap to make detailed comparisons
on a source by source basis, we identify some general trends in our
measurements of $\alpha$ and $\zeta$, and specifically the difference in
these two angles, $\sigma$ as compared to previous studies.  The values we derive for $\sigma$ are clearly
larger (ranging from 9 to 66 degrees) than in the studies of \cite{1997Effels} where they range from $-$16 to 3 degrees
at 1400 MHz and are generally less than 5 degrees, or \cite{BCW1991} where they find values for $\sigma$ from $-$14 to 12 degrees at 430 MHz and are
generally less than 5 degrees. Another contributing factor to the discrepancy in emission heights could be the result of profile evolution altering the determined values of $\phi_{\rm em}$ relative to higher frequencies.  For example, B0950+08 has an emission height that is consistent with zero and has two slightly separated components in the LWA frequency range.  At 400 MHz the two components blend together and the emission height is estimated to be $\sim$600 km by \citet{BCW1991}.

\section{Discussion and Conclusions}\label{sec:conclusion}

Using LWA1, we were able to obtain rotation measures for 15 pulsars with comparable precision to values derived from measurements at higher frequencies, and used our rotation measures to derive measurements of the local Galactic magnetic field. Our RM results show the need for improved ionospheric modeling so that better precision can be achieved at frequencies below 100 MHz.  The recent work of \citet{mal18} using {\it in situ} measurements of the ionosphere  with dual-channel GPS receivers is one avenue that may help.
RMs can also be used to calibrate ionospheric models; see, for example, \cite{Soto2013}, in which the authors test their ionosphere modeling software by observing four pulsars with LOFAR at different times of day from widely-spaced stations within the array and with the higher-frequency Westerbork Synthesis Radio Telescope. Pulsar RMs are assumed to be constant over short timescales, so observations of a pulsar with a rotation measure known precisely from low-frequency observations can show the diurnal variation in the ionosphere. 
%There is agreement between our RMs and those of \cite{LOFAR2015} and the other high-frequency observations published in the ATNF Pulsar Database.

We have presented 37 polarized profiles for 15 pulsars, and derived the {emission heights} corresponding to observing frequency, extending our knowledge of the pulsar emission region to lower frequencies and higher emission heights than previously known based on a geometric model.  

%The RVM model for calculating emission heights appears to break down at these low frequencies due to strong profile evolution.

In the future as the pulsar moves through space, precise RMs can tell us about the variation in the magnetic field of the interstellar medium along the arc of sky that the pulsar travels. Pulsars can be used to probe magnetic fields of objects other than the interstellar medium as well. Precision RMs obtained at long wavelengths can be used to find the time-varying strength of the magnetic field of some object between the observer and the pulsar, such as the Sun, as long as the effects of the ionosphere are corrected. \cite{TimHow2016} were able to use LWA1 to study a coronal mass ejection when PSR B0950+08 was about $5^\circ$  from the Sun. It has also been proposed by \cite{BU2017} to use pulsars {to study planetary environments}, although the alignment would have to be very fortunate, and planets that emit strongly in radio such as Jupiter may overwhelm the pulsar signal.

Although ionospheric effects reduce precision, low-frequency monitoring of pulsar polarimetry has great promise for future applications in studies of the interstellar medium, the heliosphere, and pulsars themselves.

\section*{Acknowledgements}

We thank J. Malins, T.J. Pearson, and an anonymous referee for constructive suggestions.

Construction of the LWA has been supported by the Office of Naval Research under Contract N00014-07-C-0147 and by the AFOSR. Support for operations and continuing development of the LWA1 is provided by the Air Force Research Laboratory and the National Science Foundation under grants AST-1835400 and AGS-1708855.
Part of this research made use of the EPN Database of Pulsar Profiles maintained by the University of Manchester, available at: http://www.epta.eu.org/epndb/

V.~Dike thanks the LSSTC Data Science Fellowship Program, which is funded by LSSTC, NSF Cybertraining Grant \#1829740, the Brinson Foundation, and the Moore Foundation; her participation in the program has benefited this work.  V.~Dike also acknowledges support from the National Science Foundation Graduate Research Fellowship Program under Grant No. DGE-1650604. 
% NOT SURE IF THIS ONE HAS TO GO HERE BUT ERRING ON THE SIDE OF CAUTION

\section*{Data Availability}

The data underlying this article are available in the LWA Data Archive
at https://lda10g.alliance.unm.edu
and can be accessed using the names of the pulsars reported in this paper.

\newpage

\bibliographystyle{mnras}
\bibliography{ms.bib} % if your bibtex file is called example.bib

\newpage

\begin{figure}
	\includegraphics[width=\columnwidth]{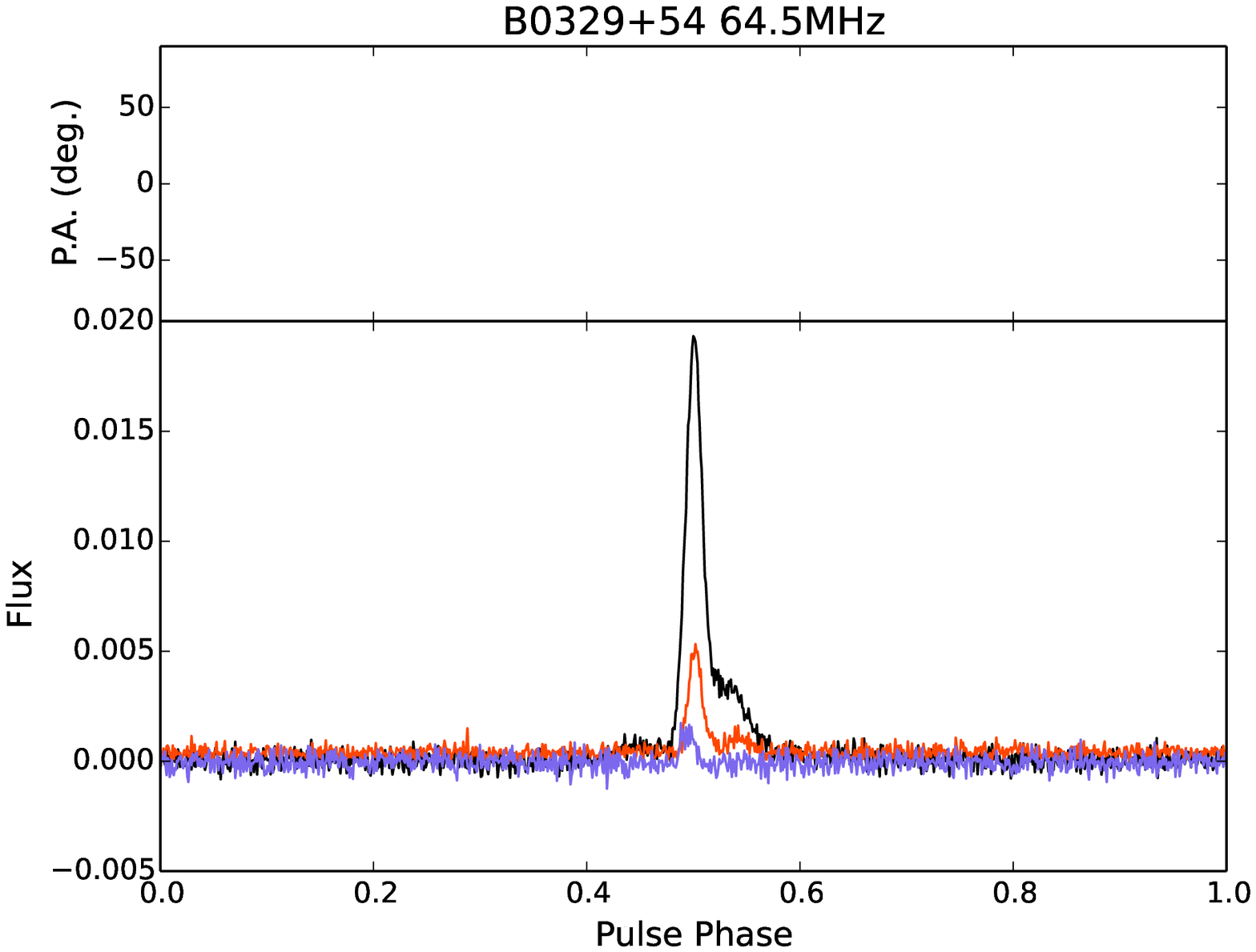}
    \includegraphics[width=\columnwidth]{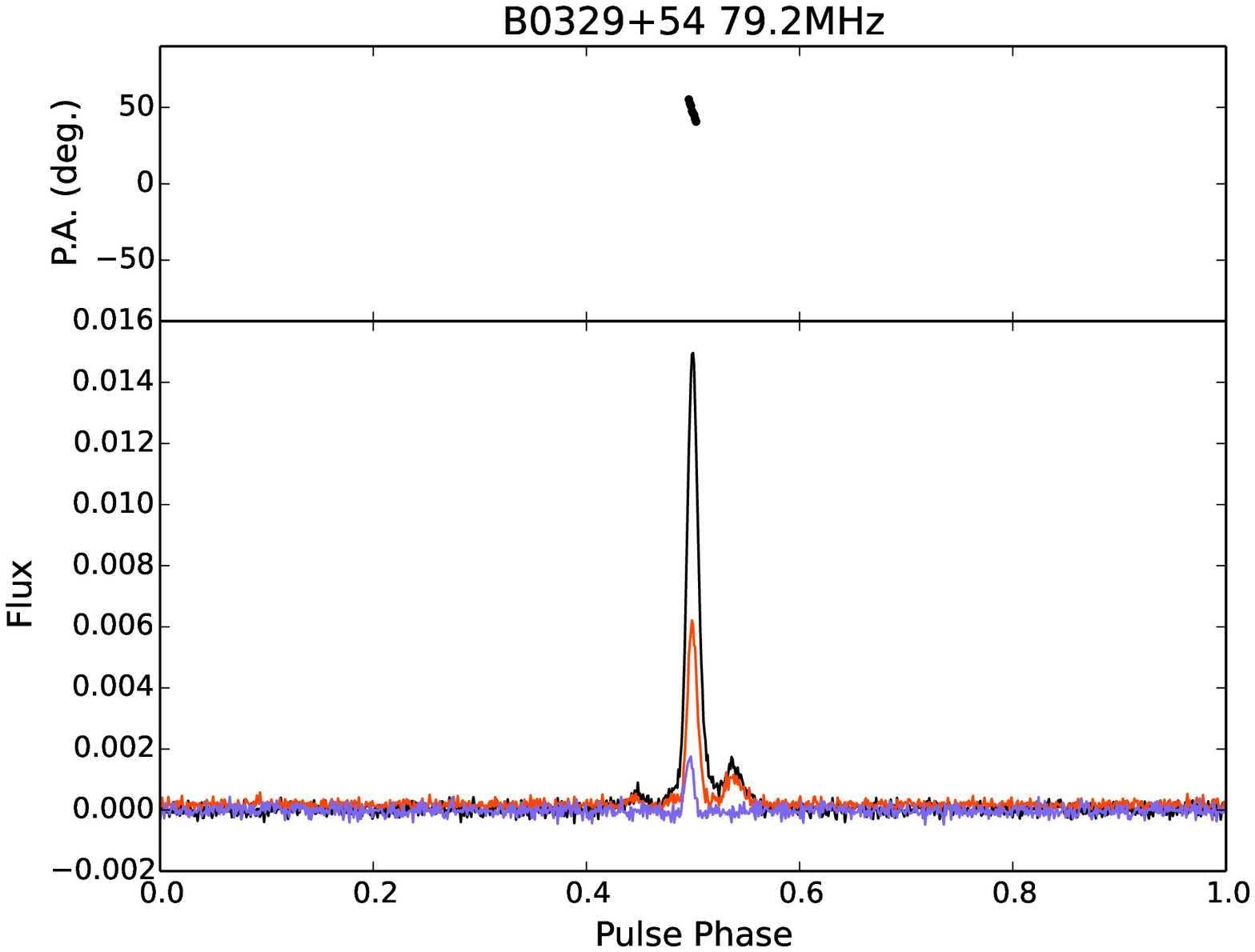}
	\caption{Profiles for B0329+54 at 64 and 79 MHz. Black, red, and blue lines correspond to total intensity, {linearly polarized intensity, and circularly polarized intensity, respectively.  The absolute flux densities have not been calibrated so the flux axis is in arbitrary units}. {The polarization position angle values over the pulse phase are shown in the top panel
	for fractional polarization values $P/I>0.3$.}
	\label{B0329fig}}
\end{figure}

\begin{figure}
	\includegraphics[width=\columnwidth]{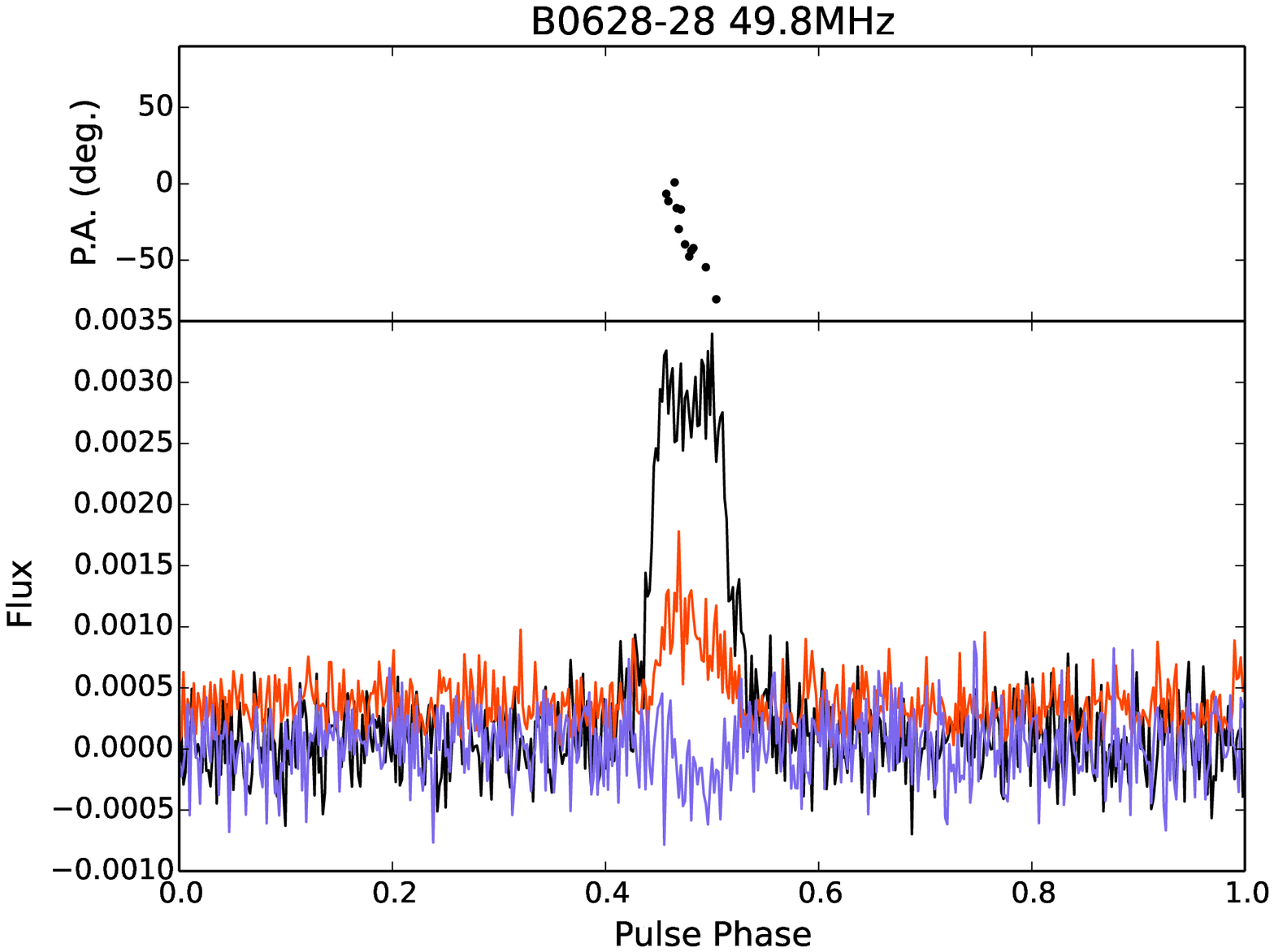}
    \includegraphics[width=\columnwidth]{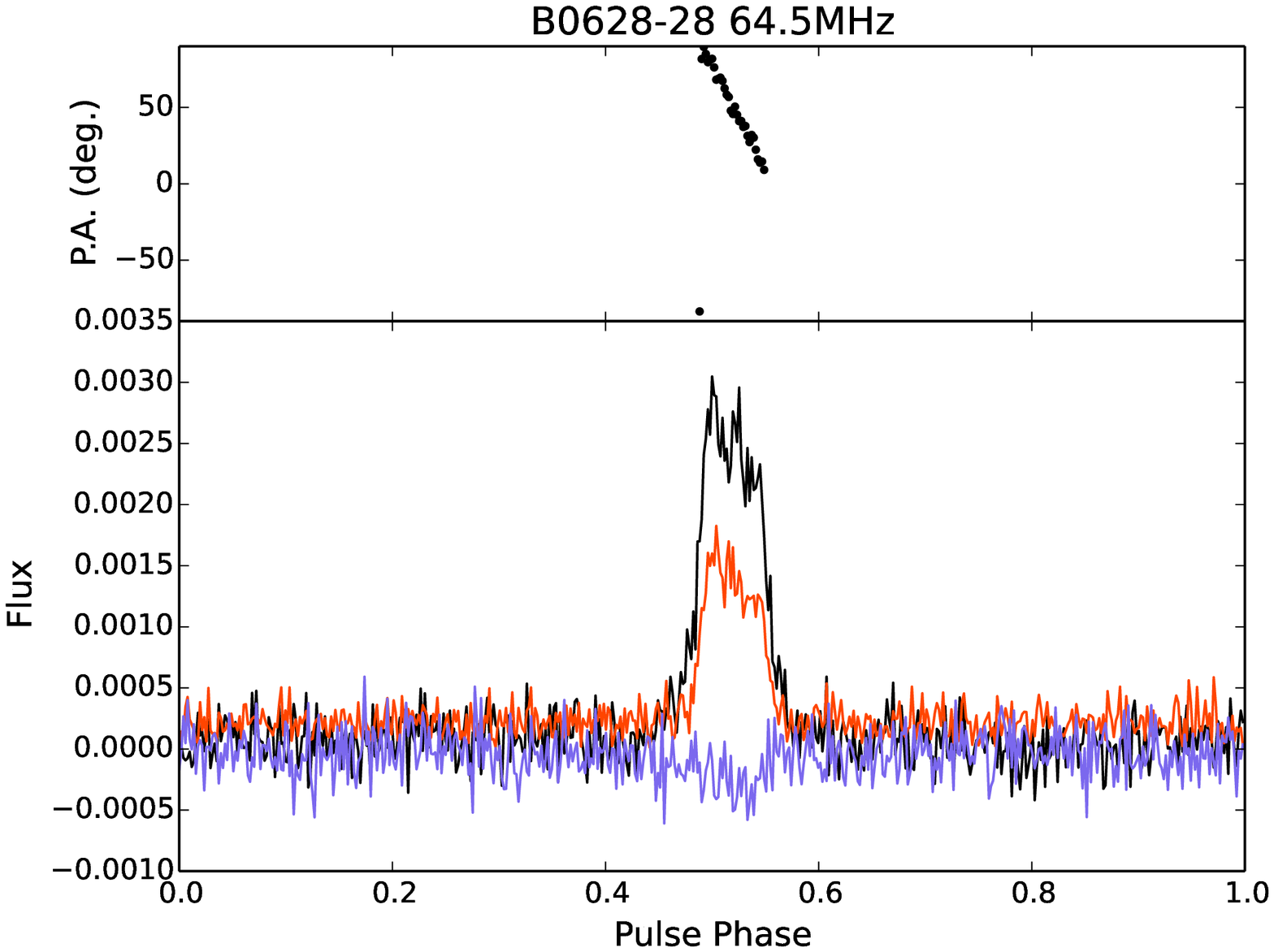}
      \includegraphics[width=\columnwidth]{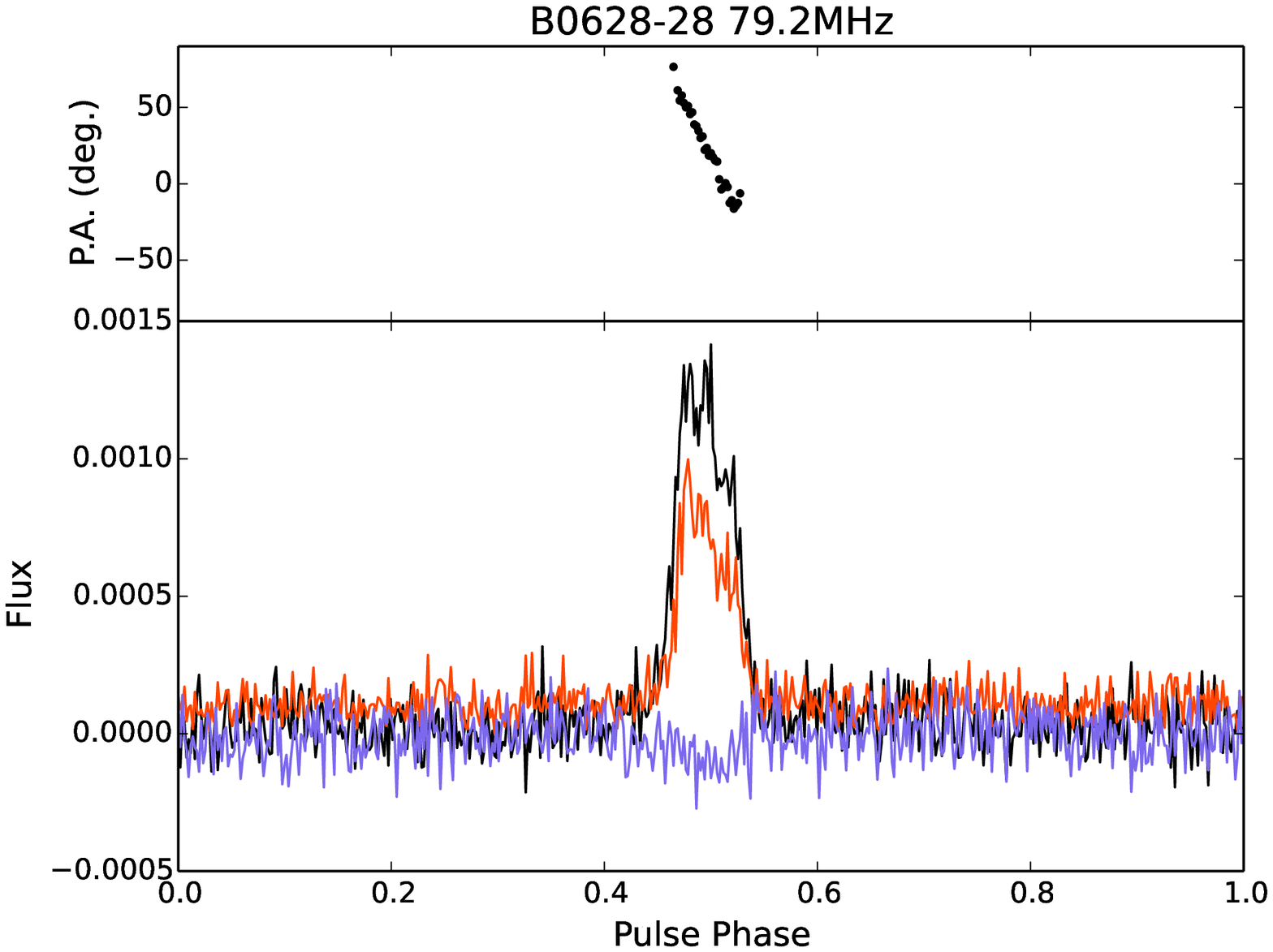}
	\caption{Profiles for B0628-28 at 49, 64 and 79 MHz with colors as described in Fig.~ \ref{B0329fig}\label{B0628fig}}
\end{figure}

\begin{figure}
	\includegraphics[width=\columnwidth]{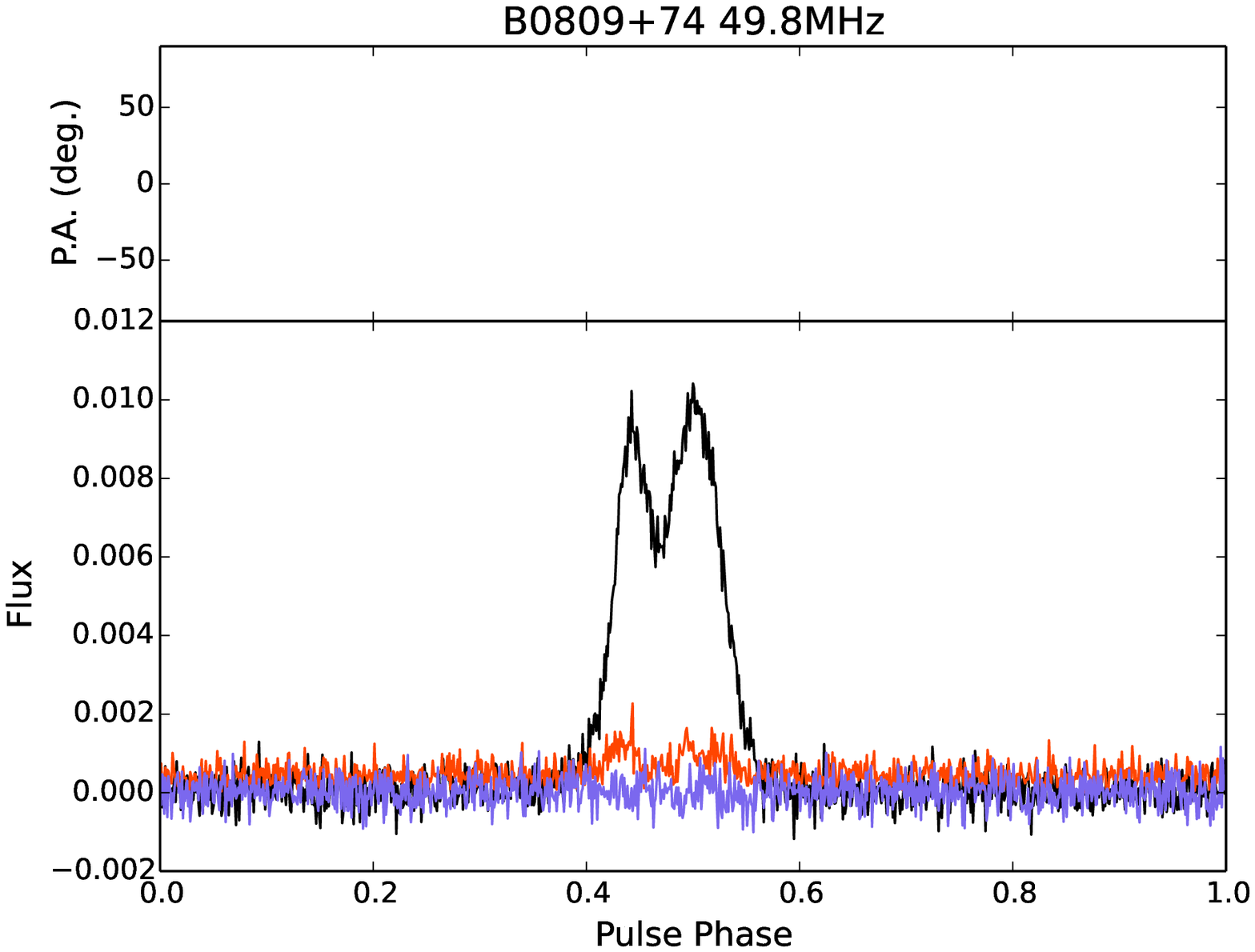}
    \includegraphics[width=\columnwidth]{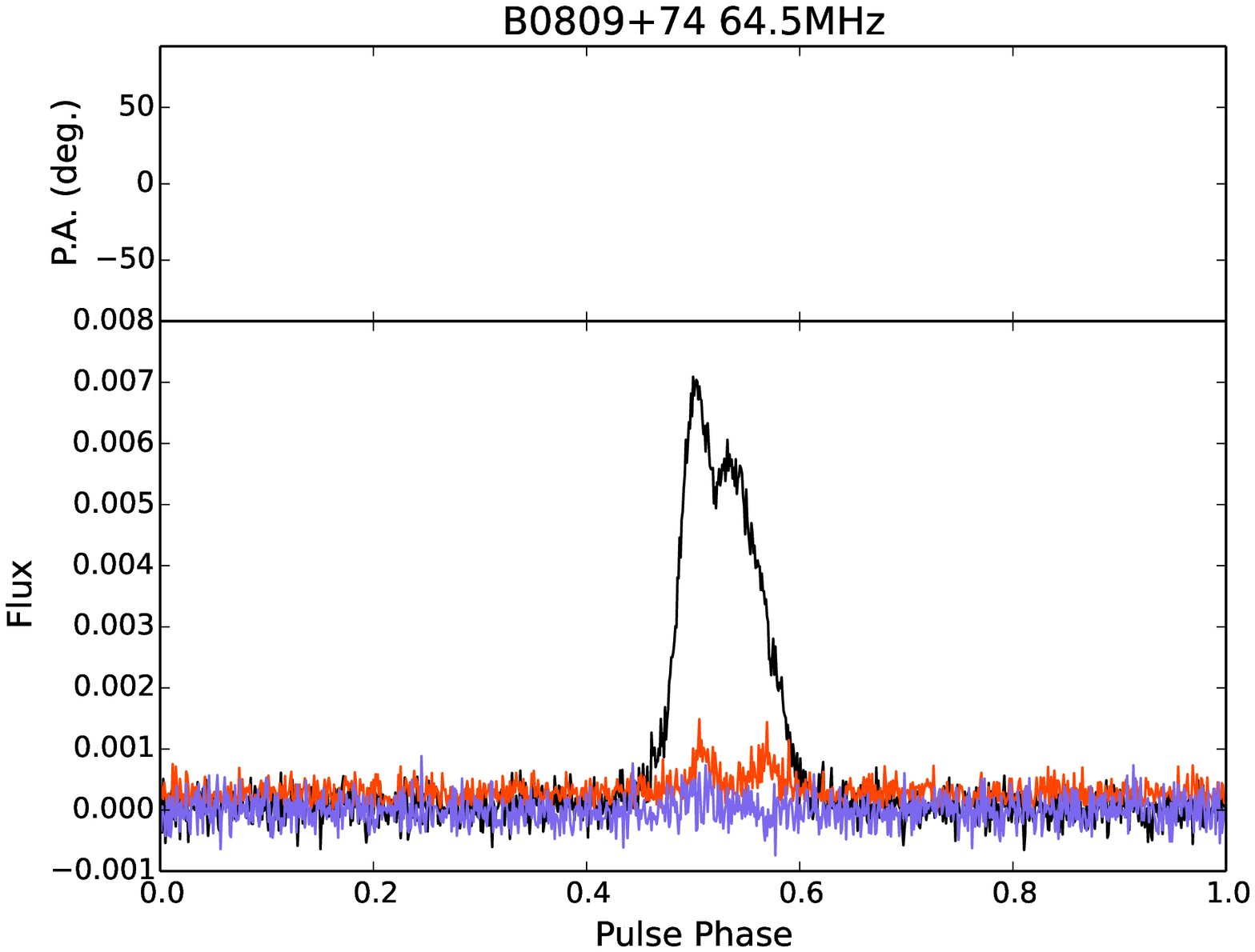}
      \includegraphics[width=\columnwidth]{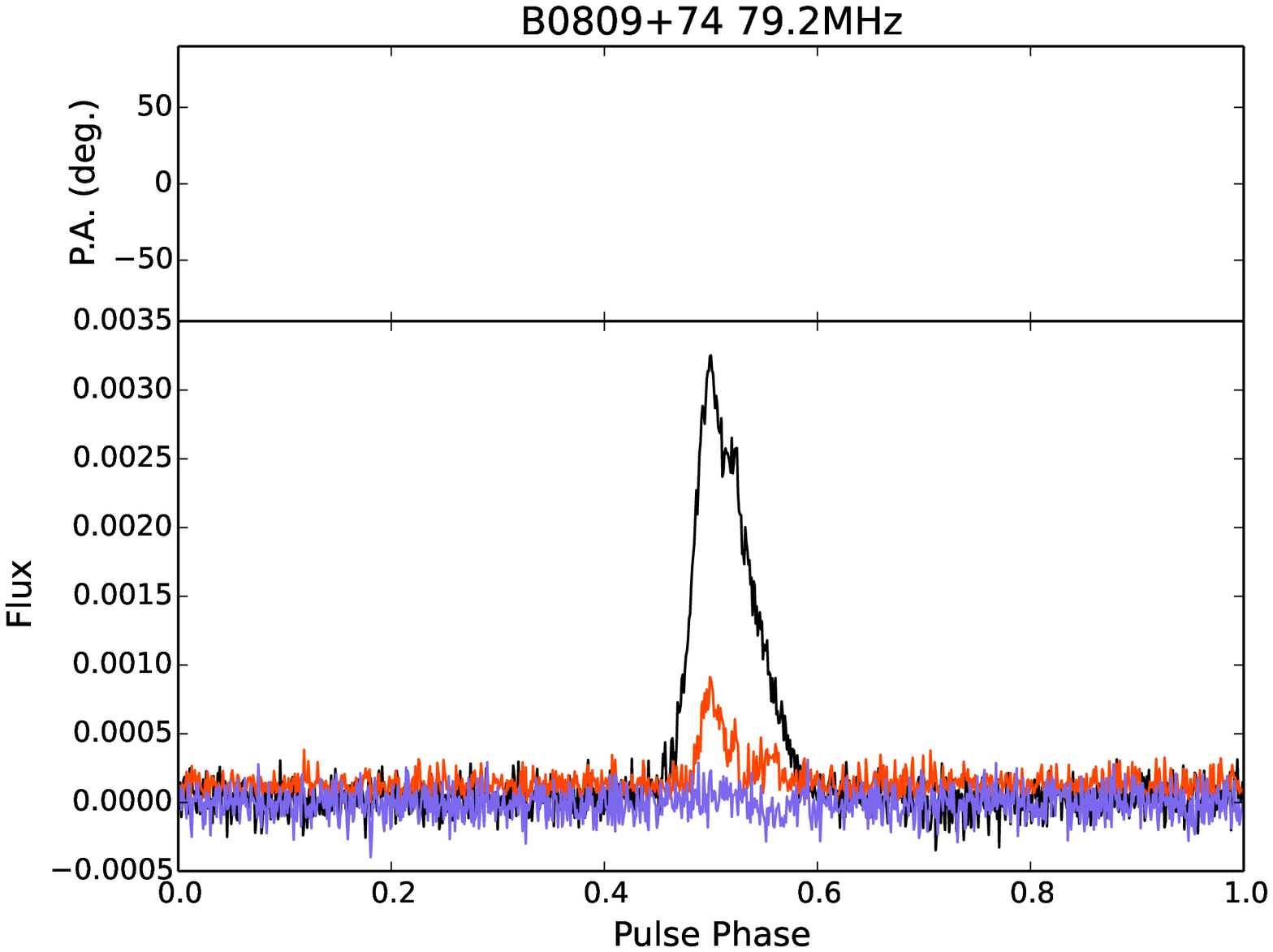}
	\caption{Profiles for B0809+74 at 49, 64 and 79 MHz with colors as described in Fig.~ \ref{B0329fig}\label{B0809fig}}
\end{figure}

\begin{figure}
	\includegraphics[width=\columnwidth]{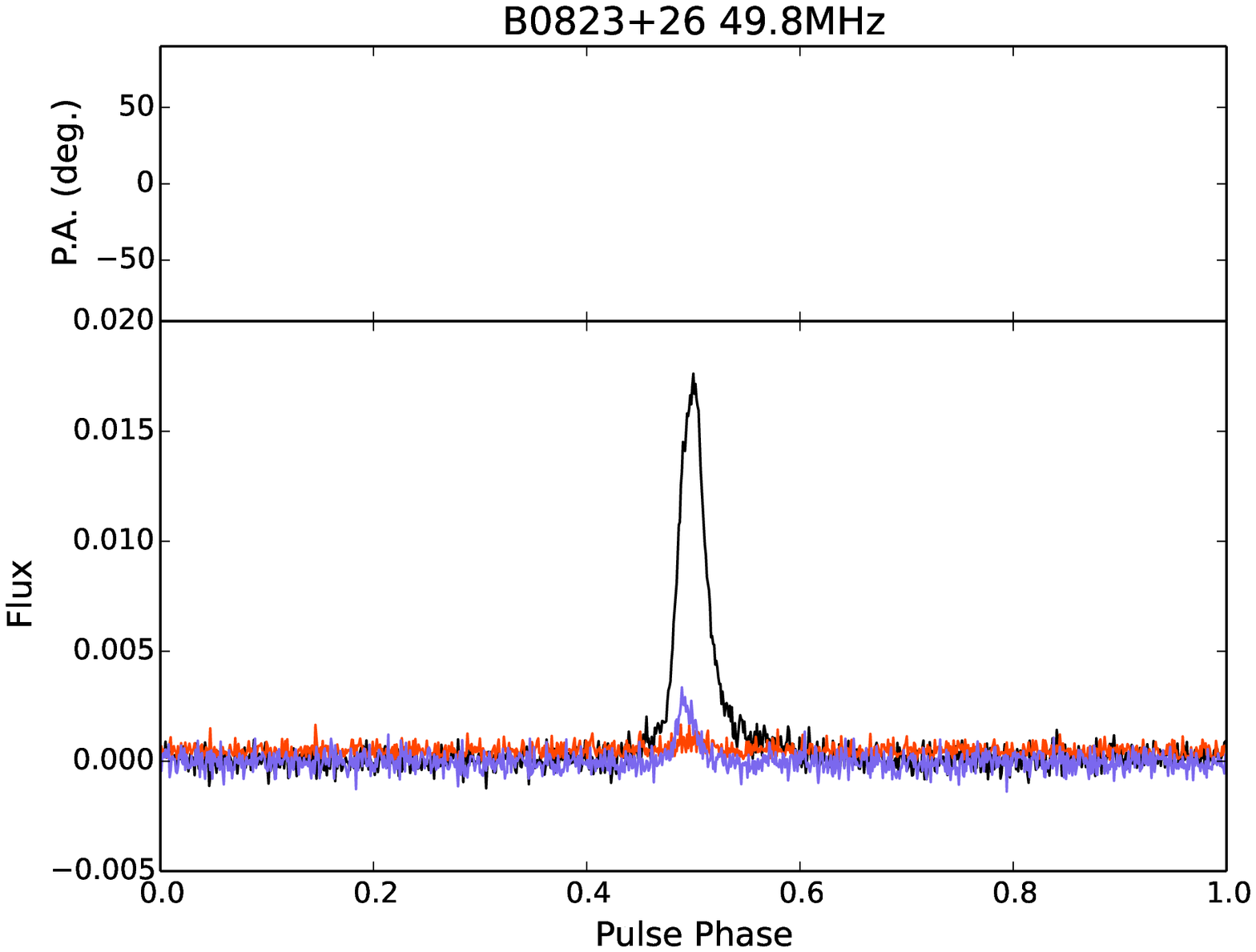}
    \includegraphics[width=\columnwidth]{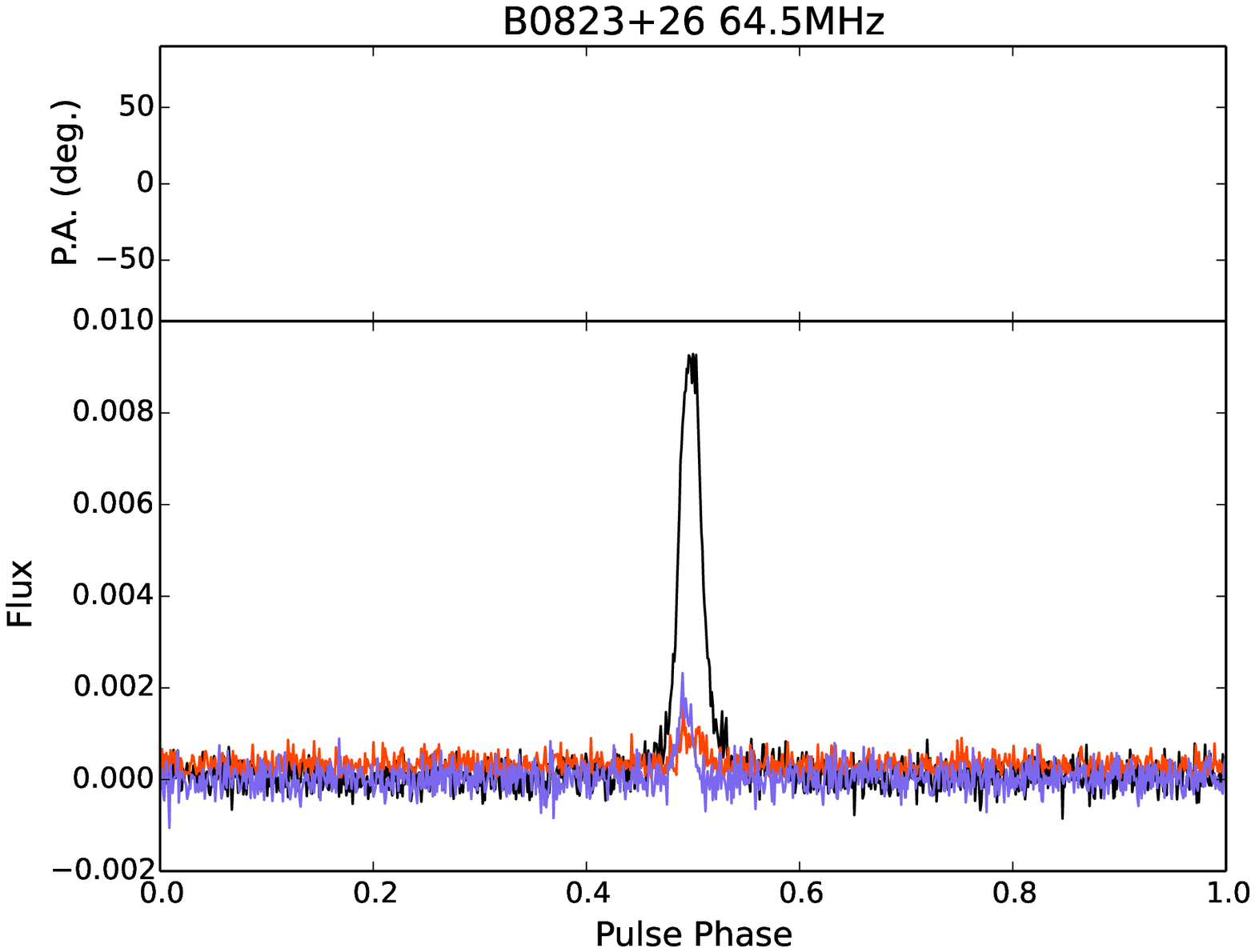}
      \includegraphics[width=\columnwidth]{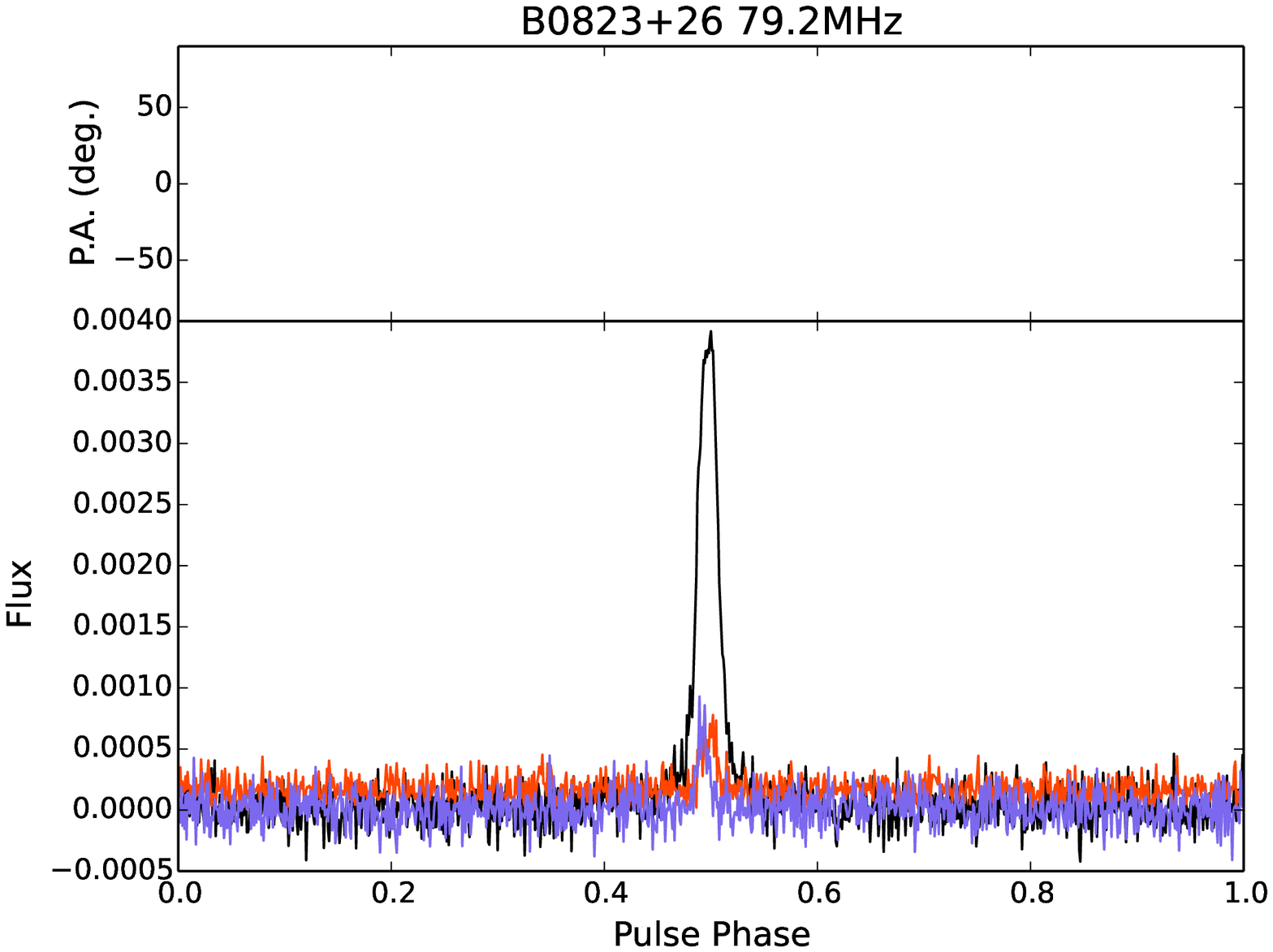}
	\caption{Profiles for B0823+26 at 49, 64 and 79 MHz with colors as described in Fig.~ \ref{B0329fig}\label{B0823fig}}
\end{figure}

\begin{figure}
	\includegraphics[width=\columnwidth]{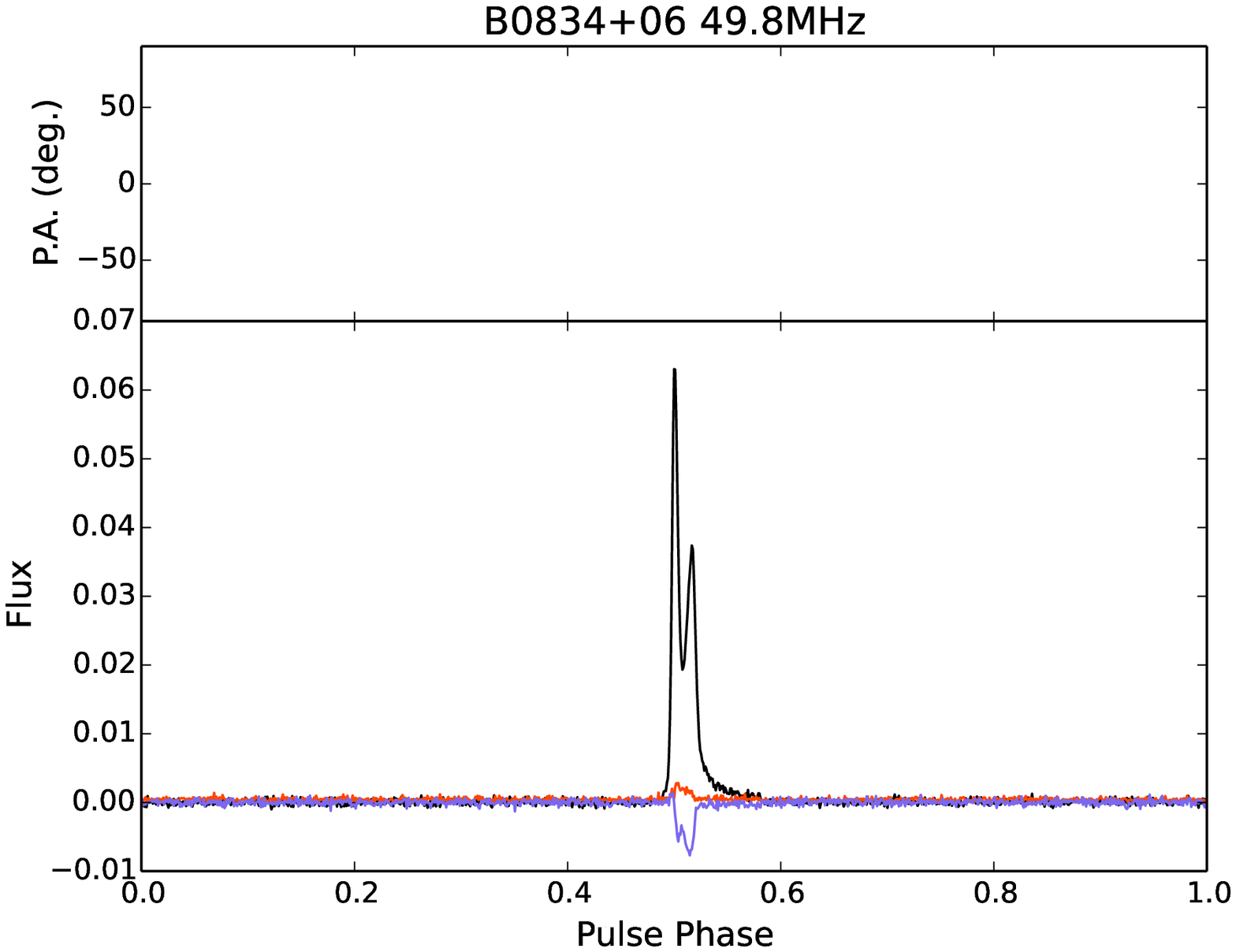}
    \includegraphics[width=\columnwidth]{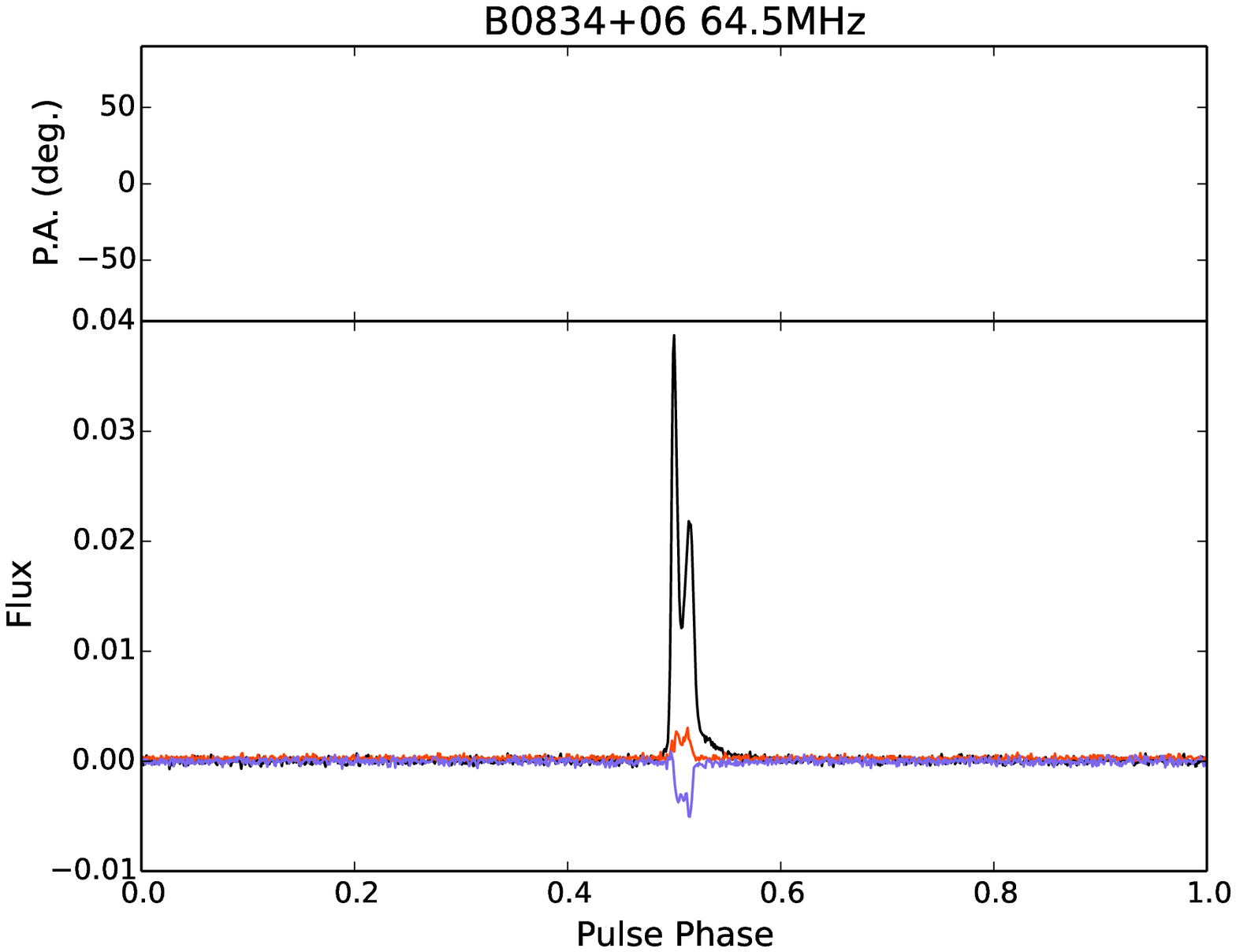}
      \includegraphics[width=\columnwidth]{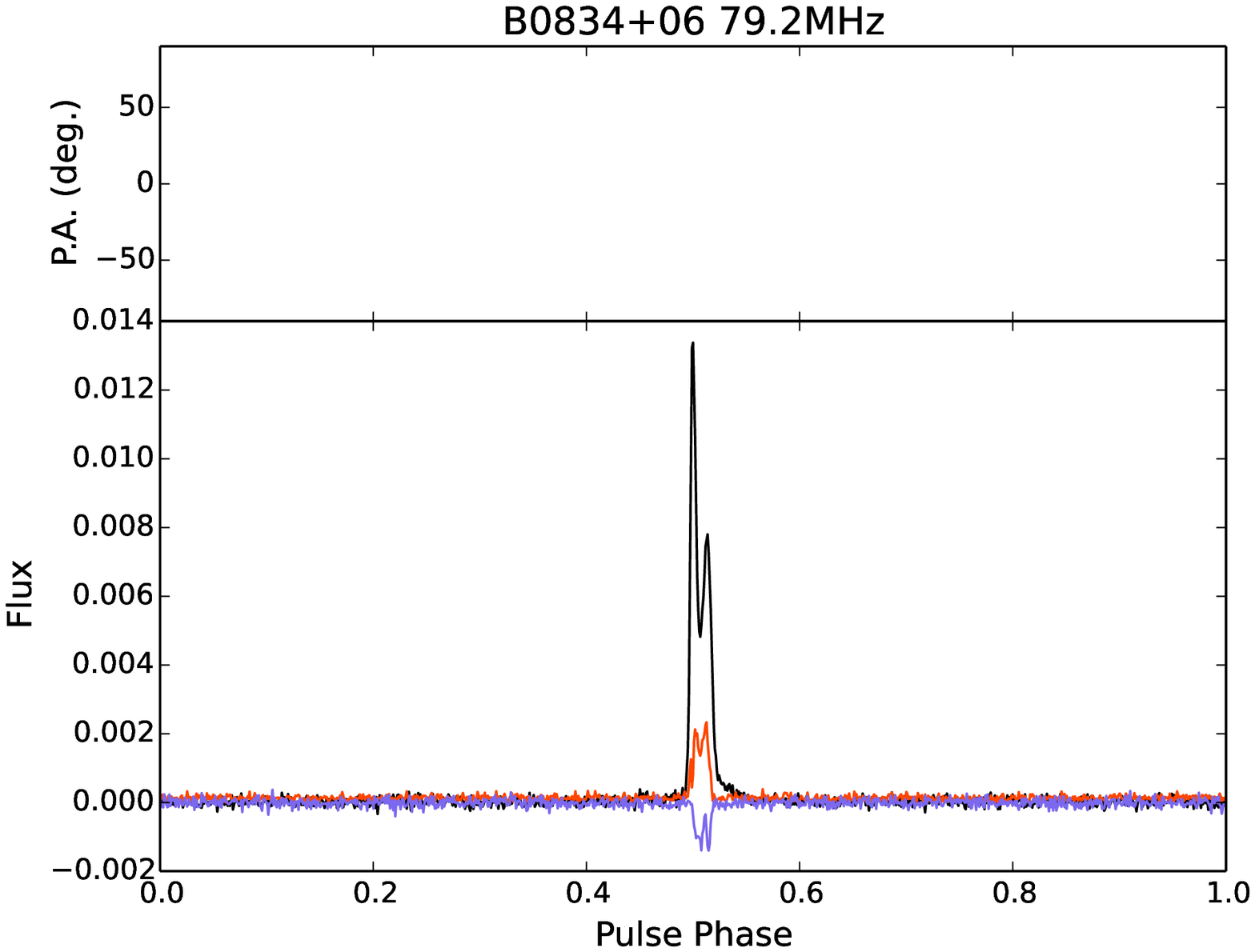}
	\caption{Profiles for B0834+06 at 49, 64 and 79 MHz with colors as described in Fig.~ \ref{B0329fig}\label{B0834fig}}
\end{figure}

\begin{figure}
	\includegraphics[width=\columnwidth]{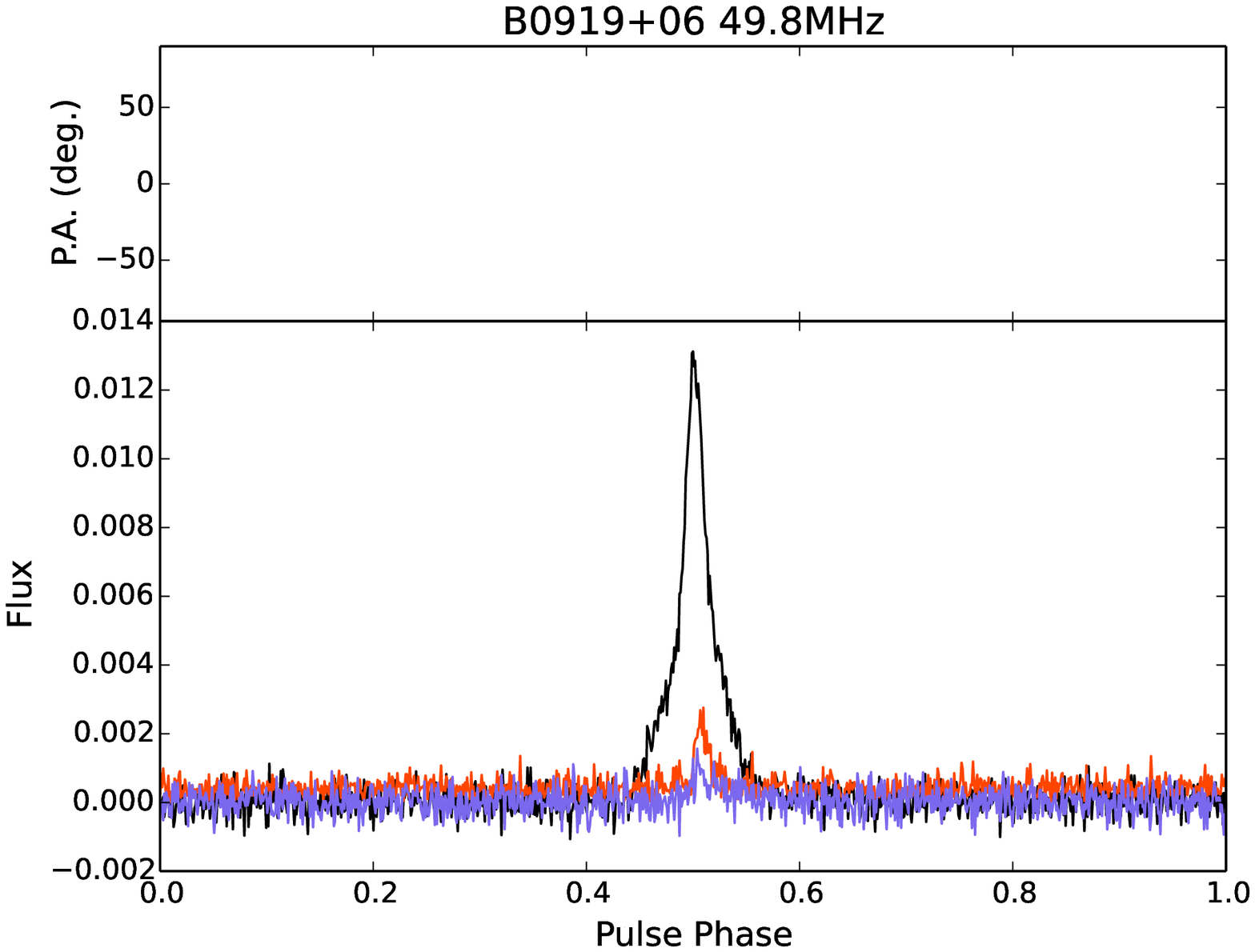}
    \includegraphics[width=\columnwidth]{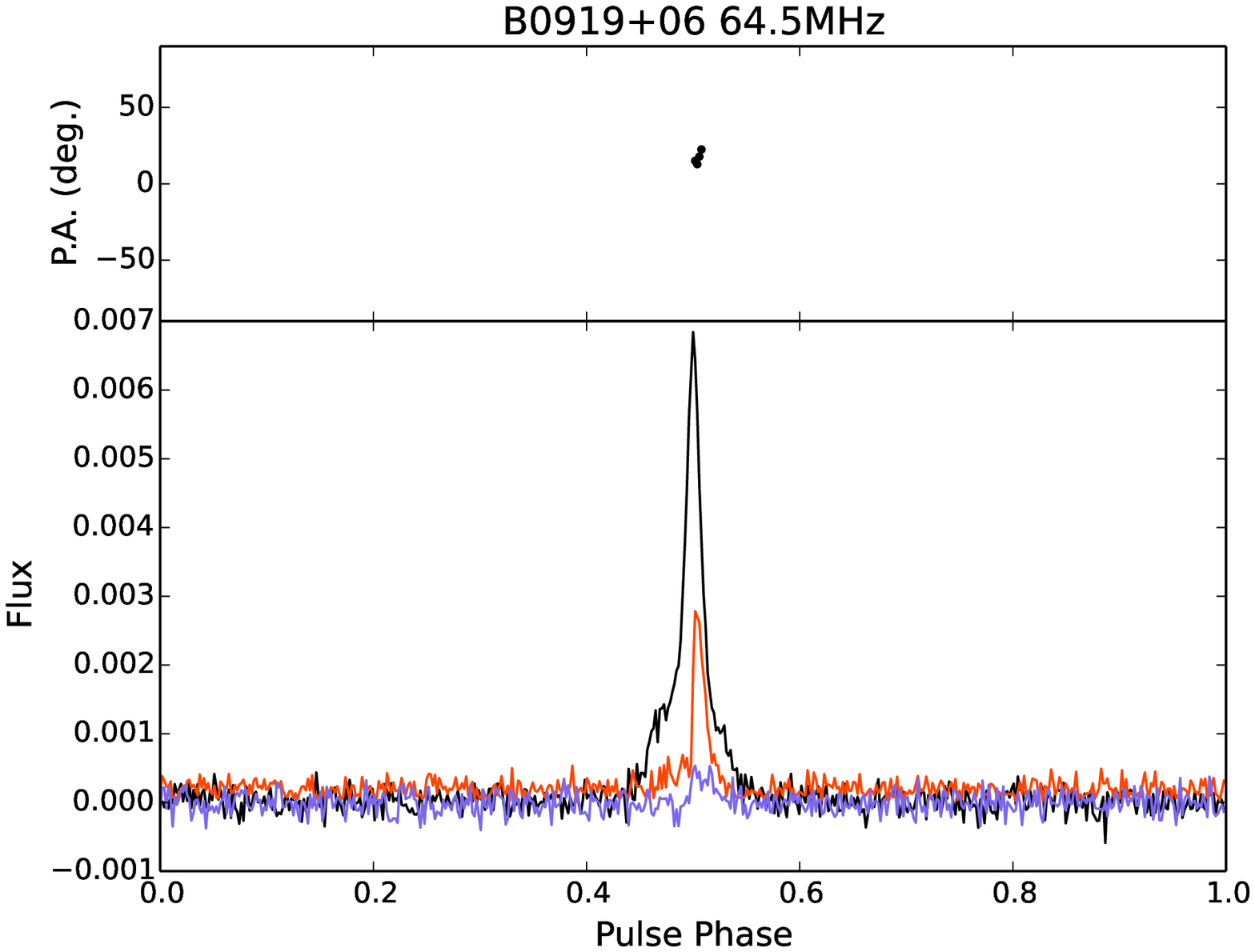}
      \includegraphics[width=\columnwidth]{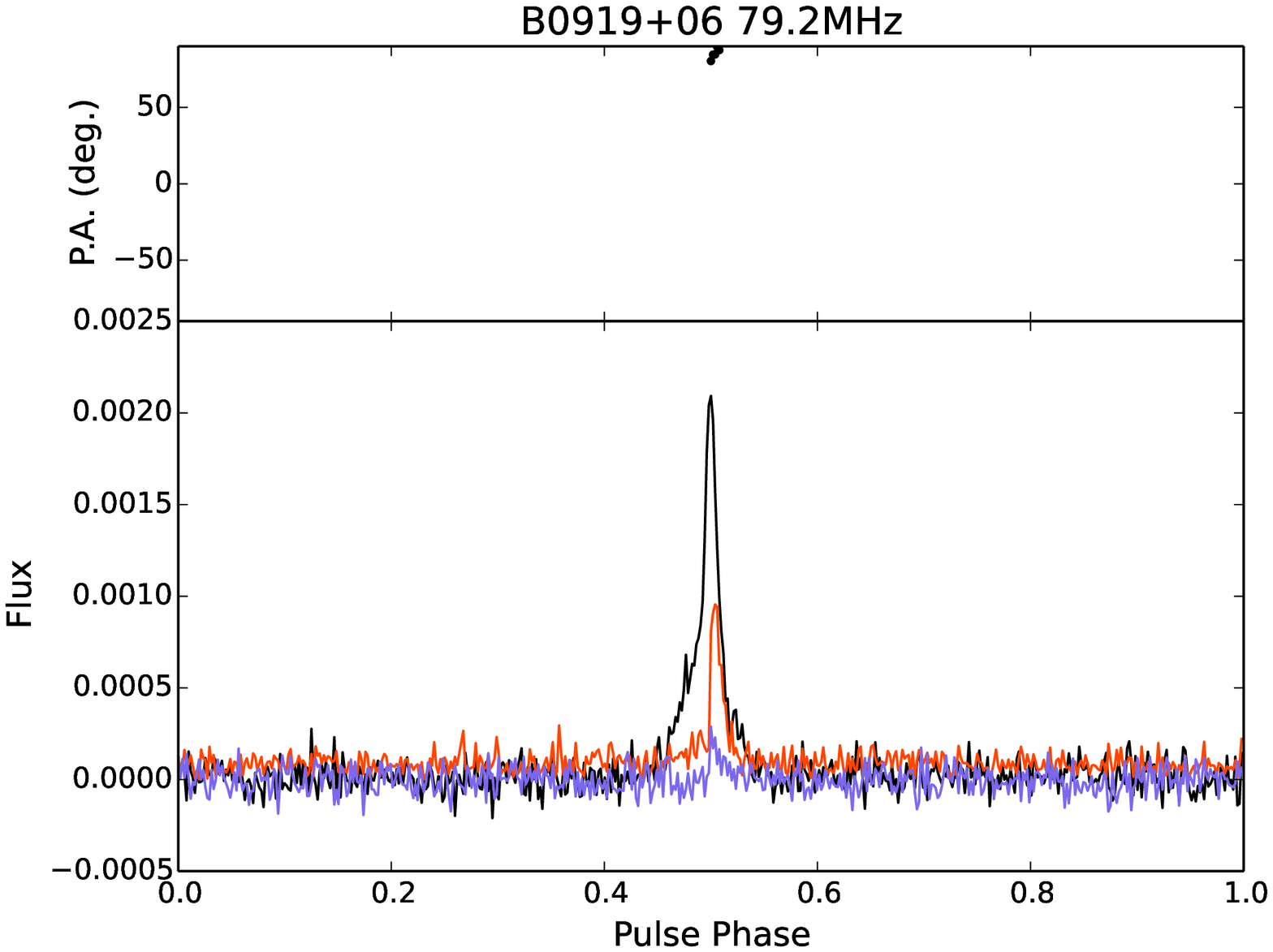}
	\caption{Profiles for B0919+06 at 49, 64 and 79 MHz with colors as described in Fig.~ \ref{B0329fig}\label{B0919fig}}
\end{figure}

\begin{figure}
	\includegraphics[width=\columnwidth]{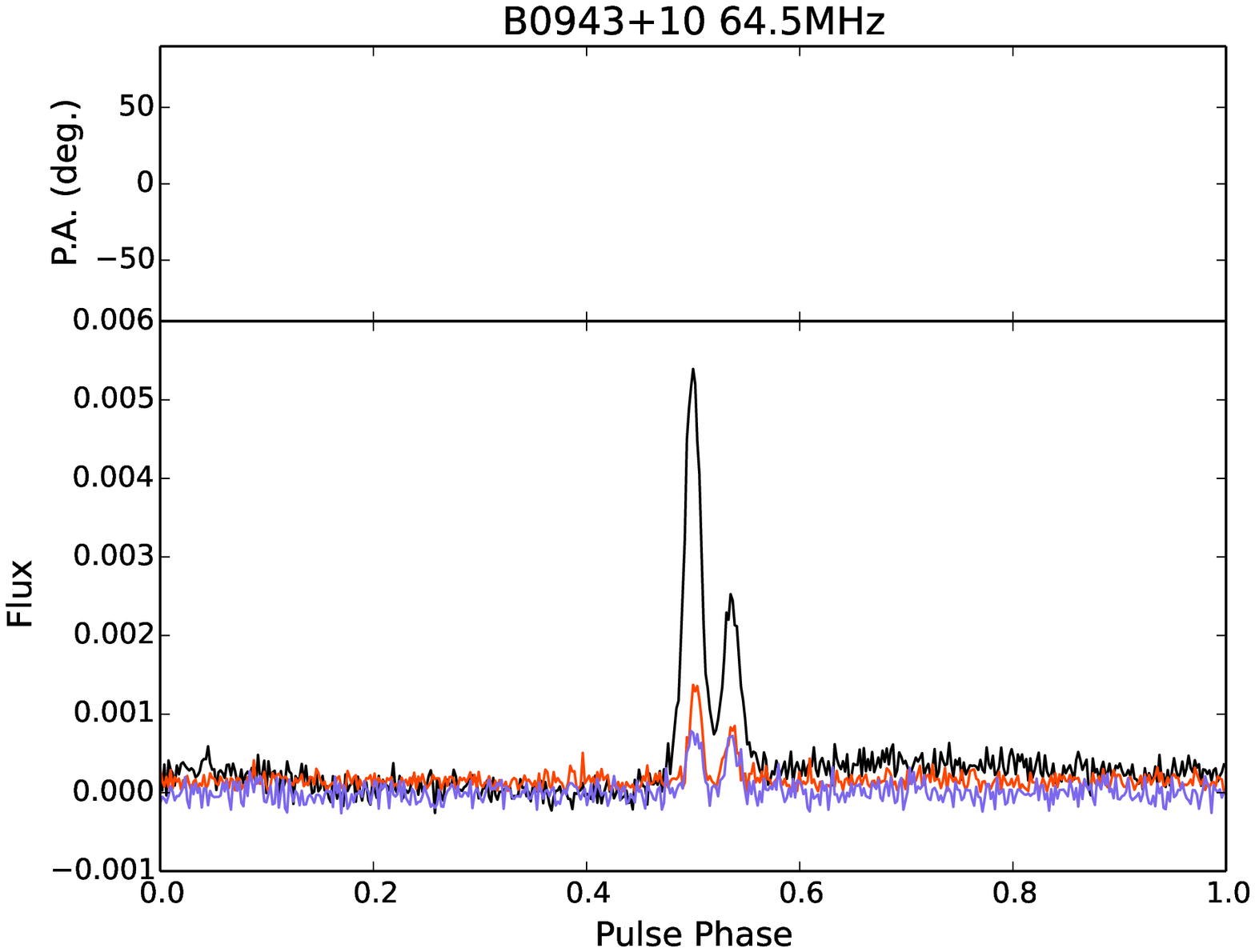}
    \includegraphics[width=\columnwidth]{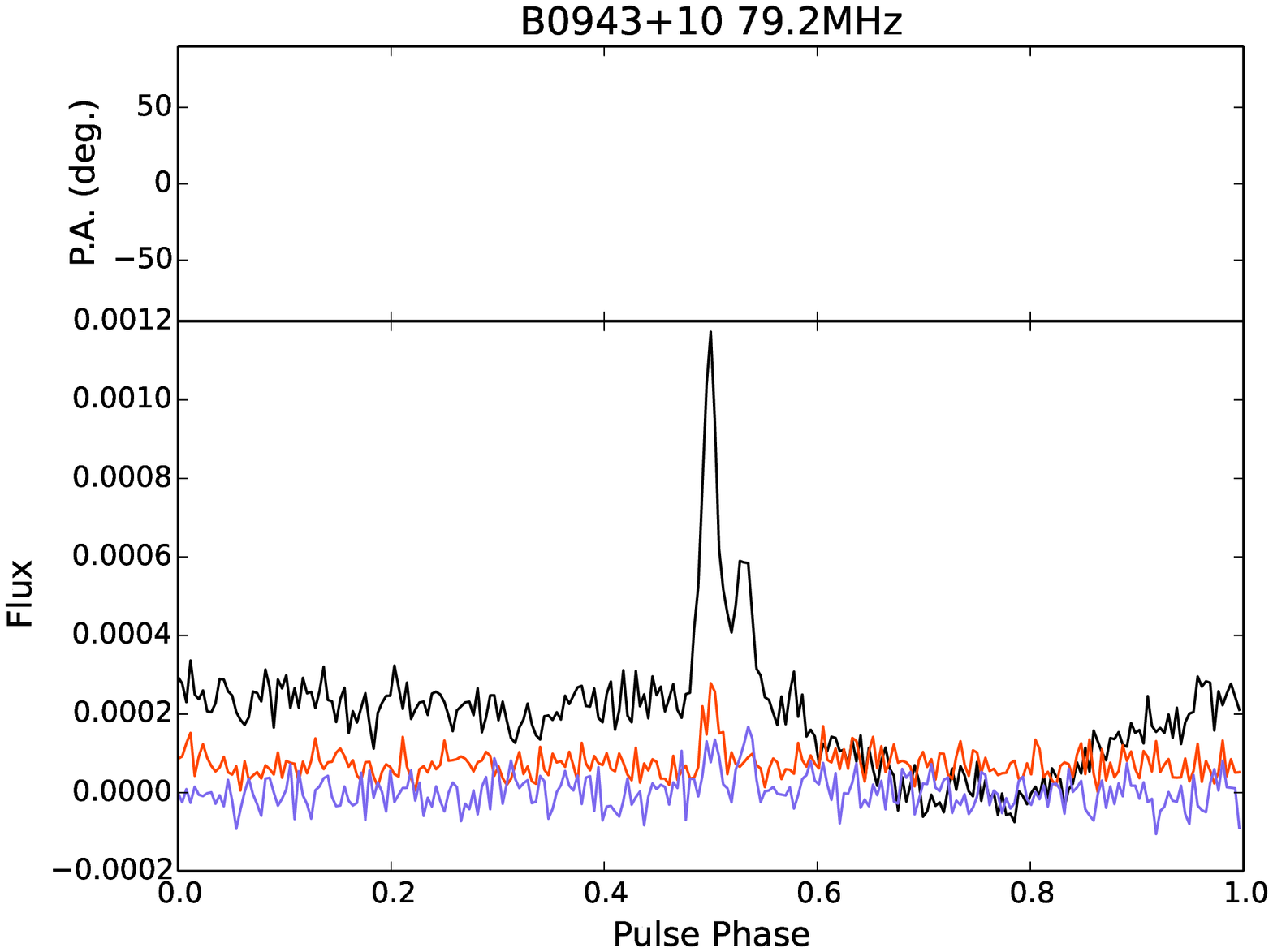}
	\caption{Profiles for B0943+10 at 64 and 79 MHz with colors as described in Fig.~ \ref{B0329fig}\label{B0943fig}}
\end{figure}

\begin{figure}
	\includegraphics[width=\columnwidth]{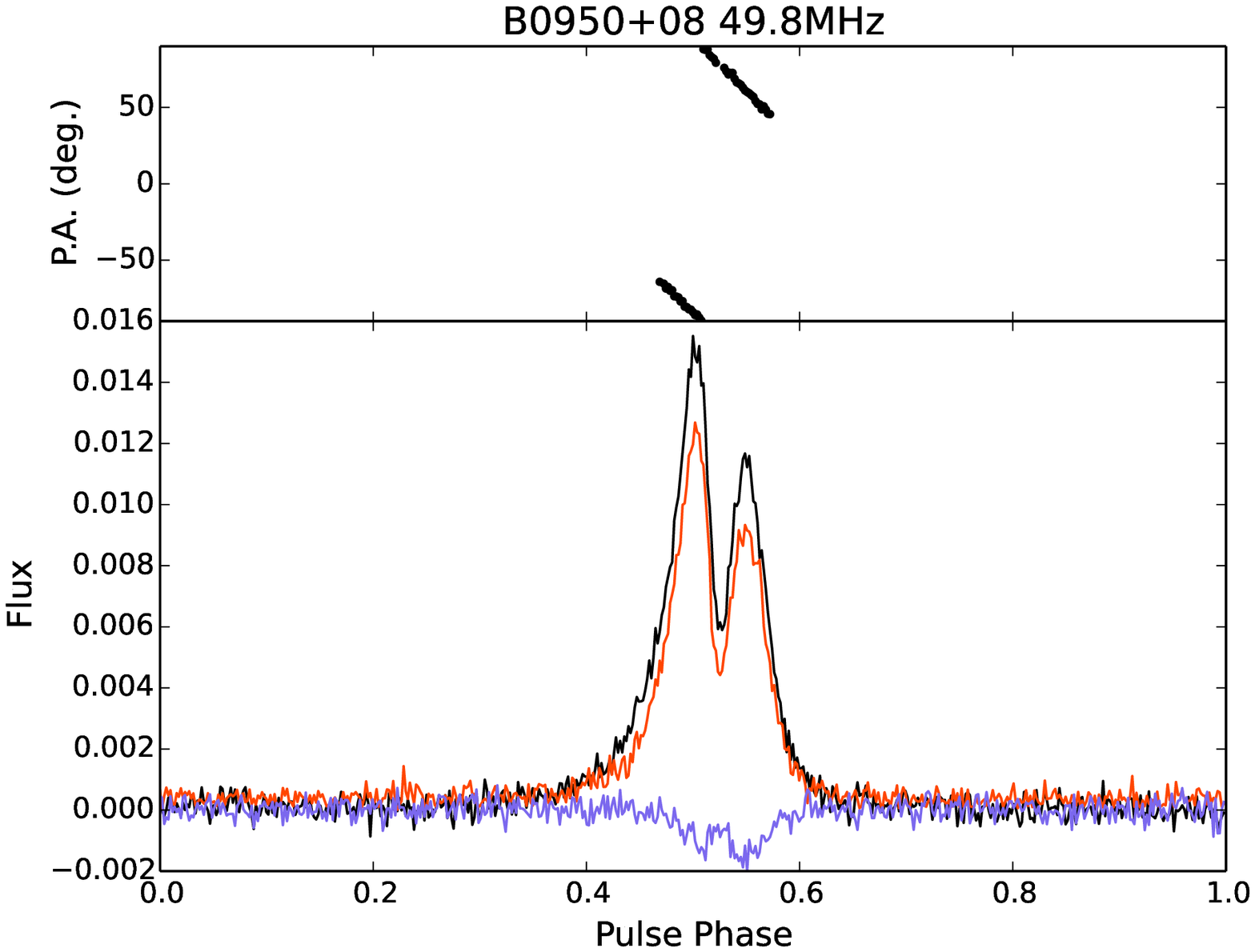}
    \includegraphics[width=\columnwidth]{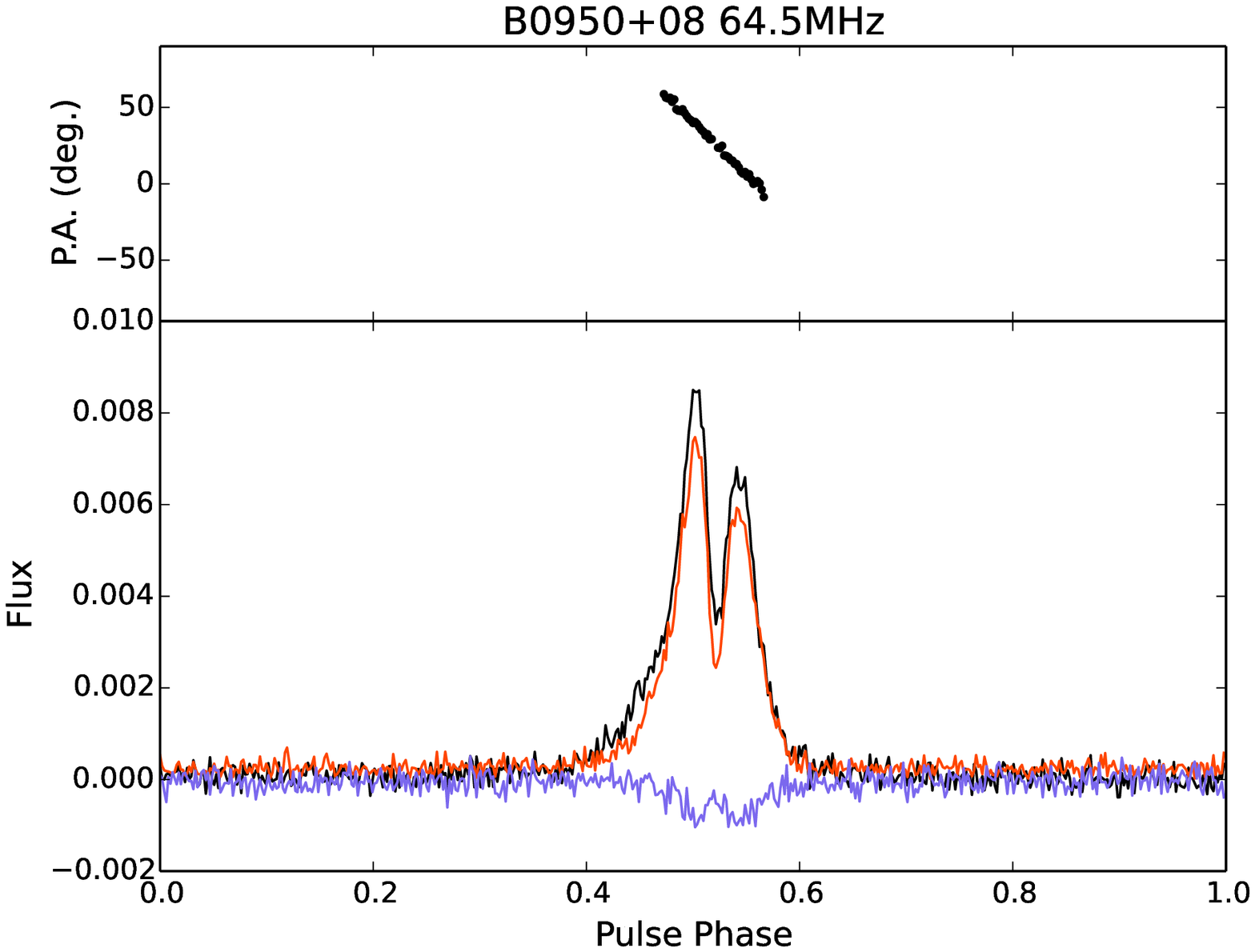}
      \includegraphics[width=\columnwidth]{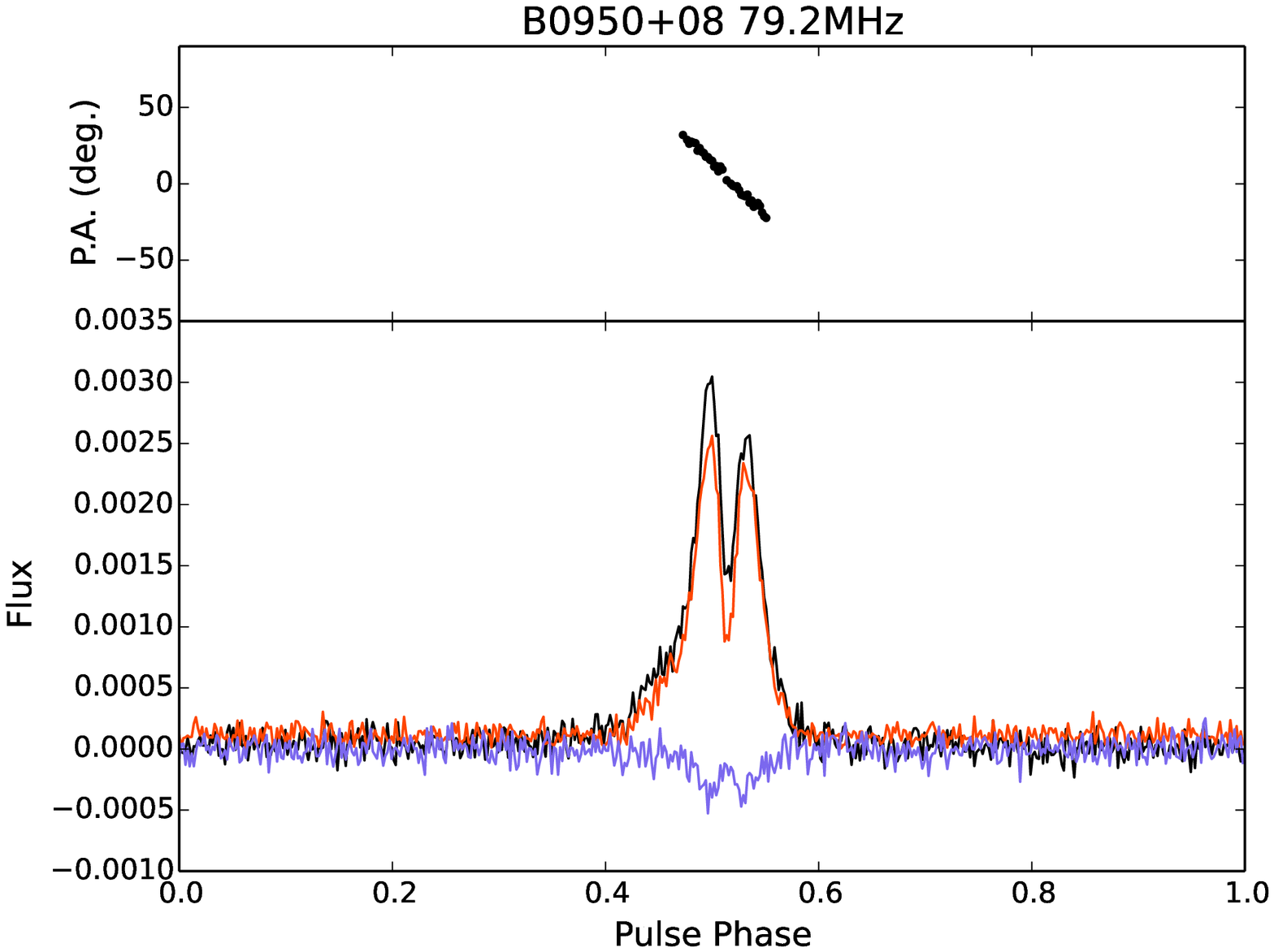}
	\caption{Profiles for B0950+08 at 49, 64 and 79 MHz with colors as described in Fig.~ \ref{B0329fig}\label{B0950fig}}
\end{figure}

\begin{figure}
	\includegraphics[width=\columnwidth]{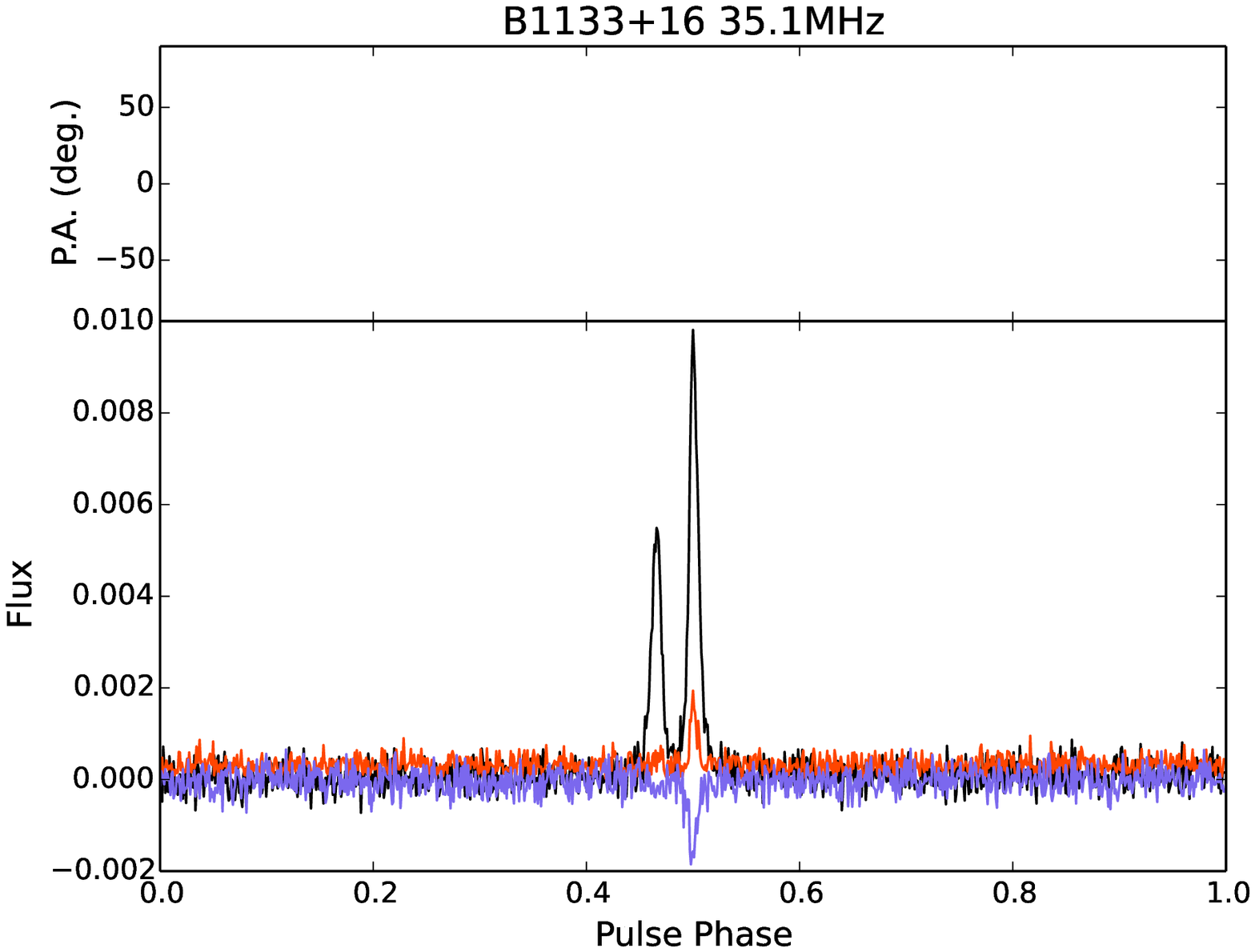}
    \includegraphics[width=\columnwidth]{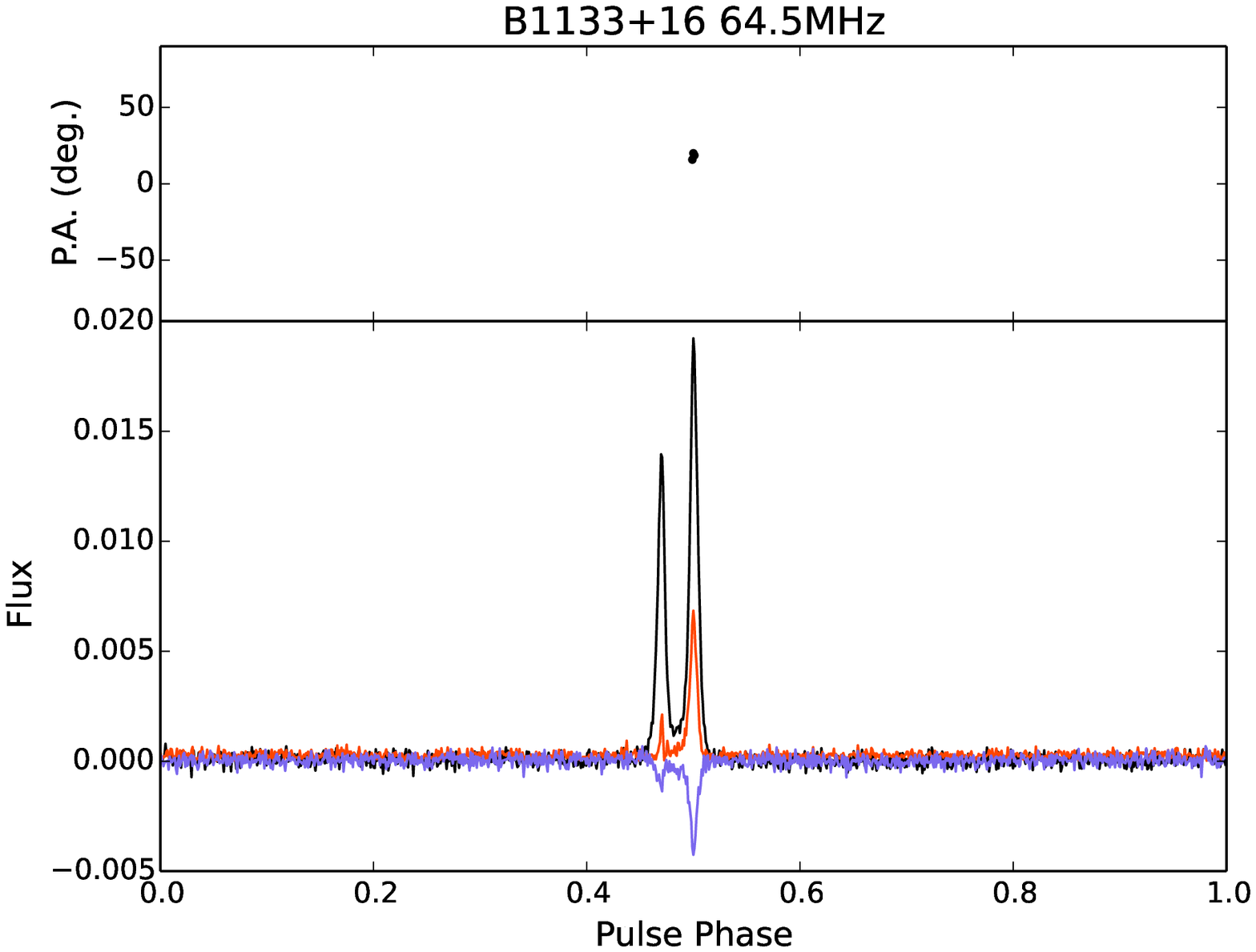}
    \includegraphics[width=\columnwidth]{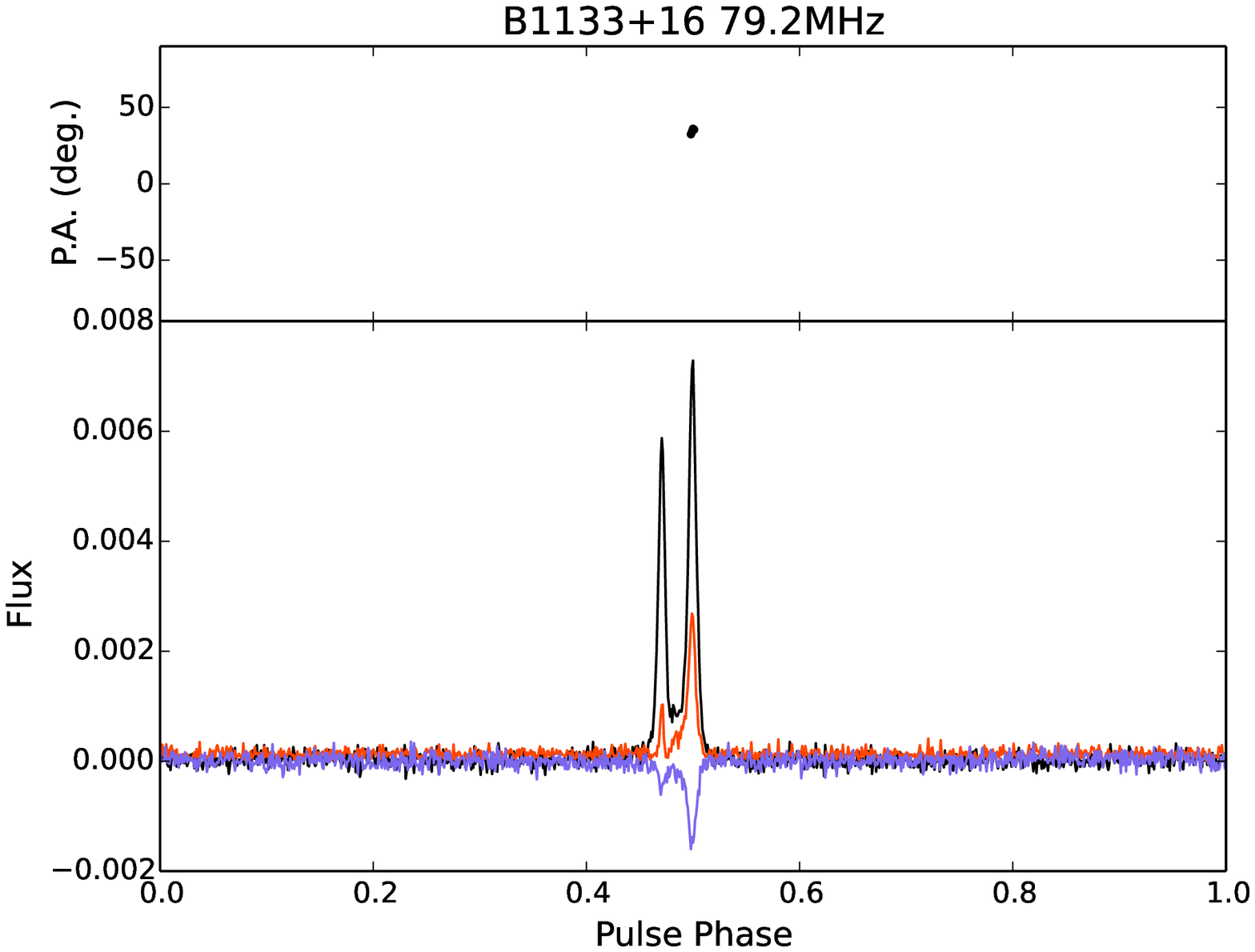}
	\caption{Profiles for B1133+16 at 35, 64 and 79 MHz with colors as described in Fig.~ \ref{B0329fig}\label{B1133fig}}
\end{figure}

\begin{figure}
	\includegraphics[width=\columnwidth]{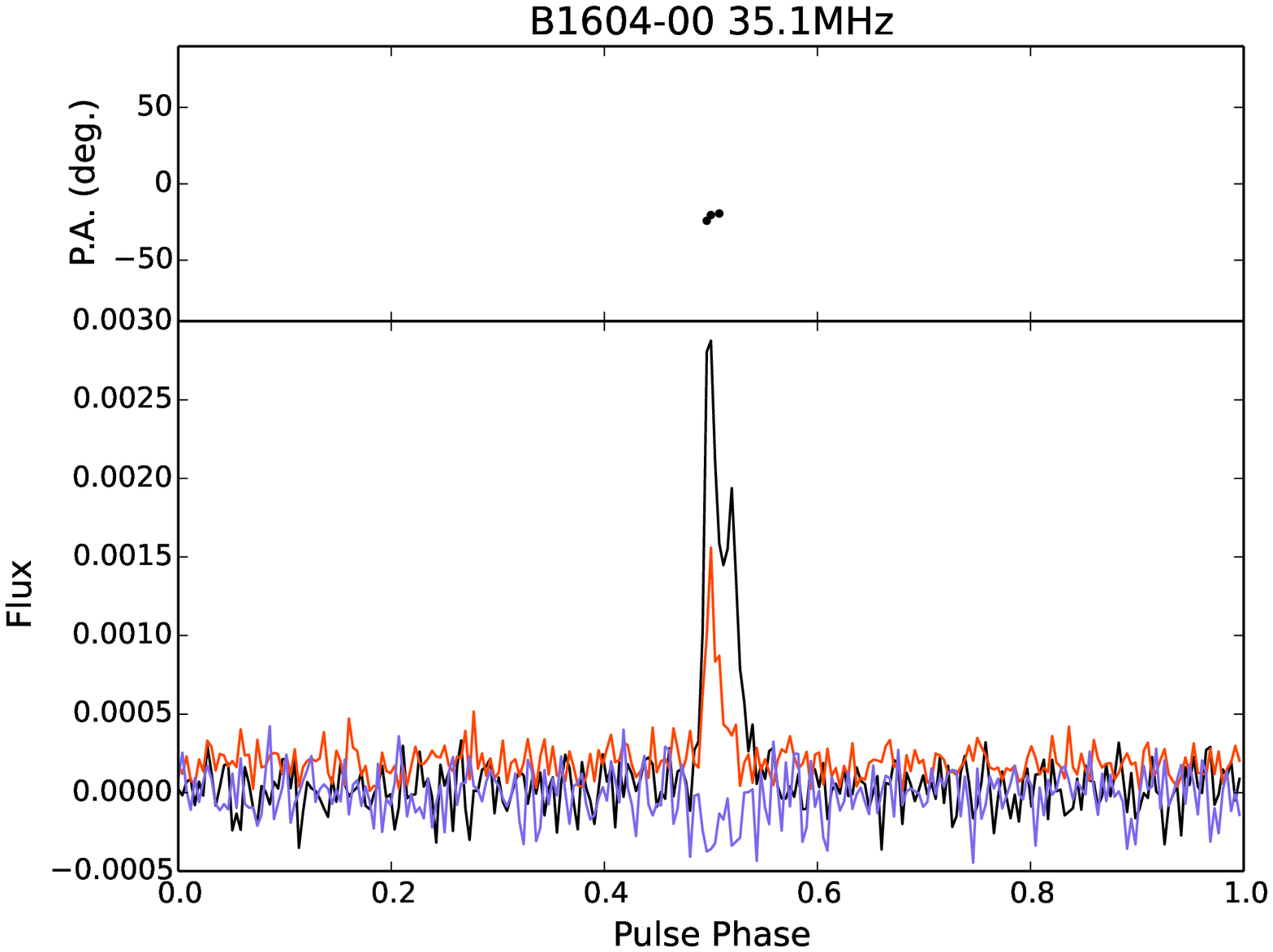}
    \includegraphics[width=\columnwidth]{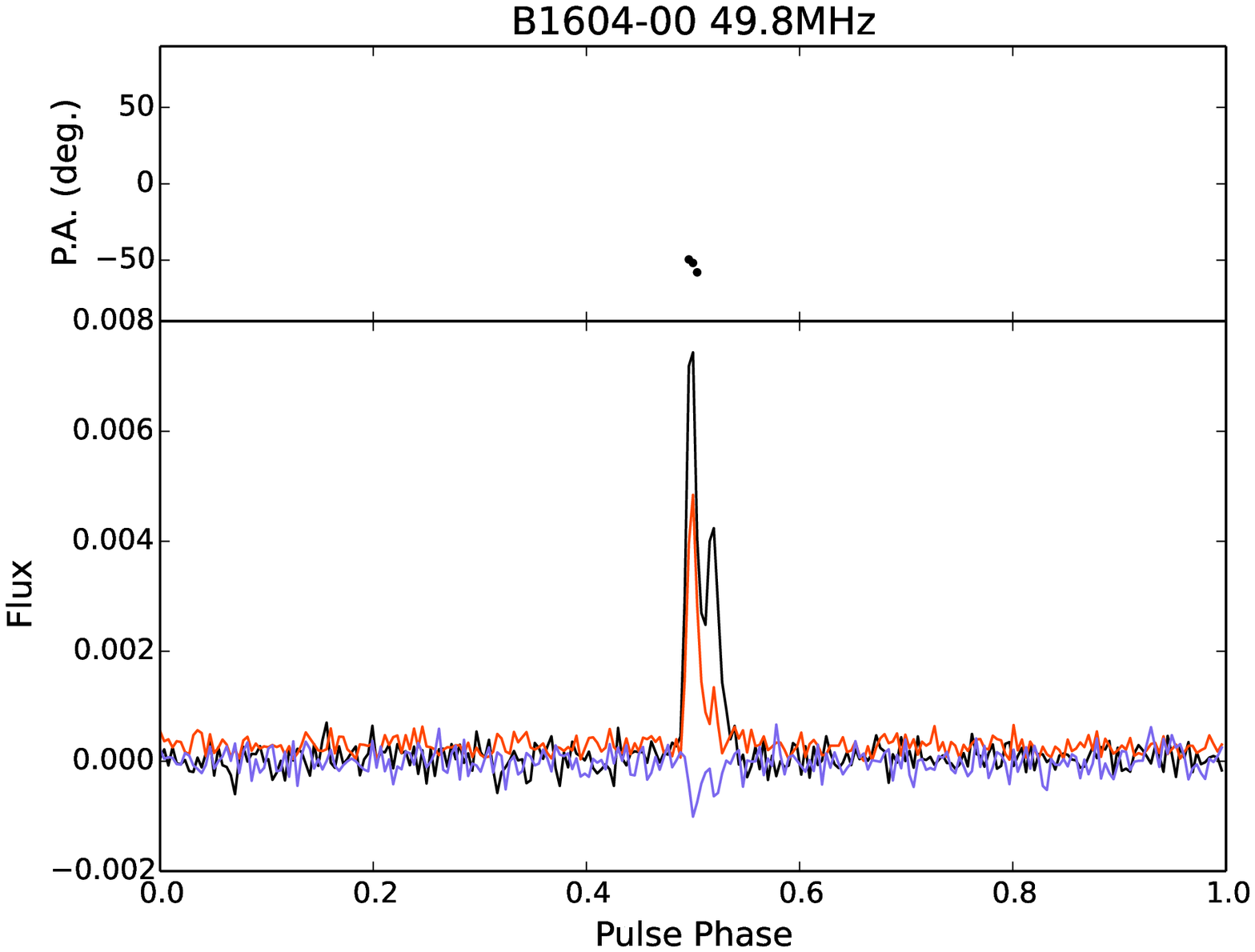}
    \includegraphics[width=\columnwidth]{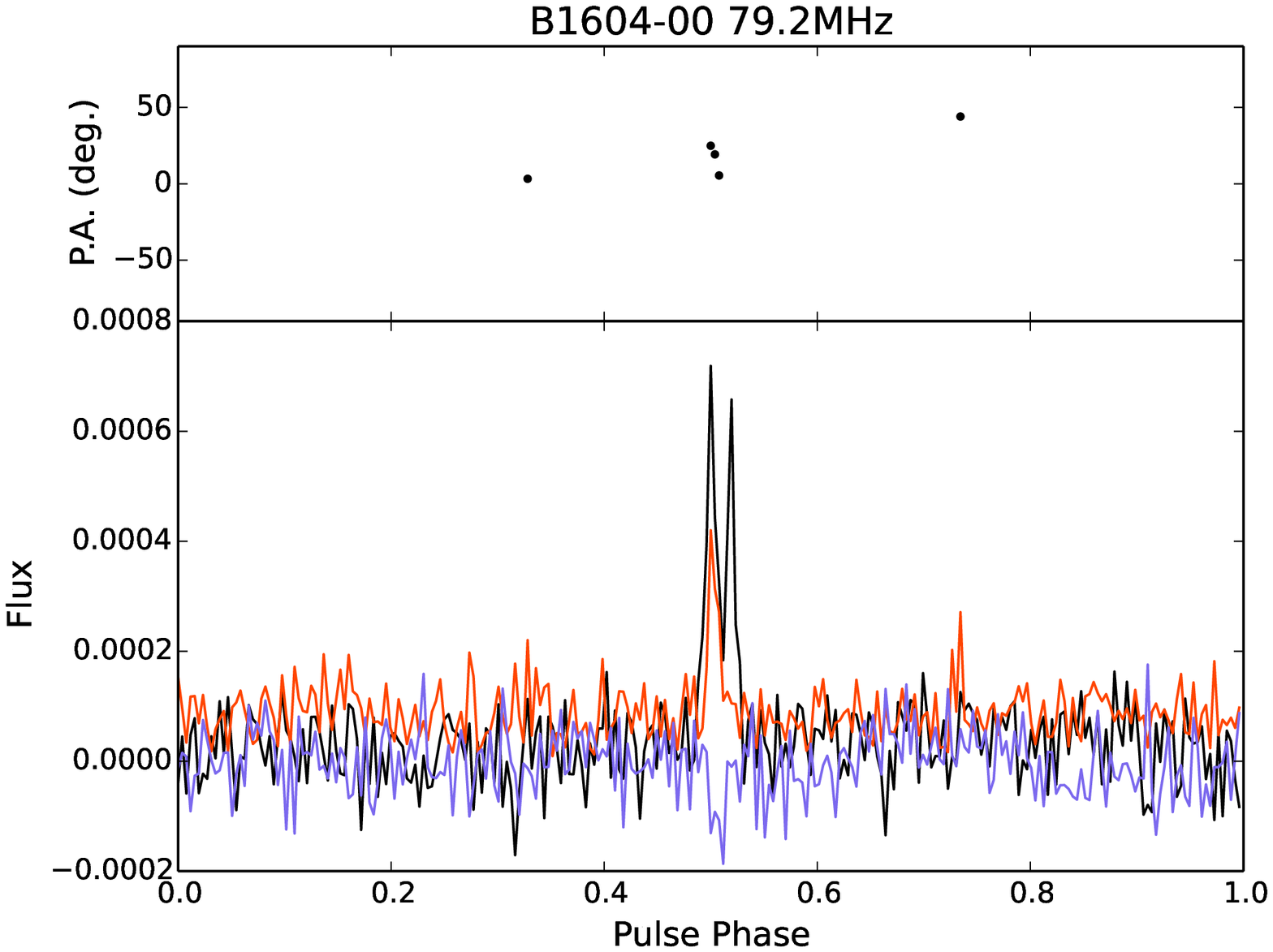}
	\caption{Profiles for B1604$-$00 at 35, 64 and 79 MHz with colors as described in Fig.~ \ref{B0329fig}\label{B1604fig}}
\end{figure}

\begin{figure}
	\includegraphics[width=\columnwidth]{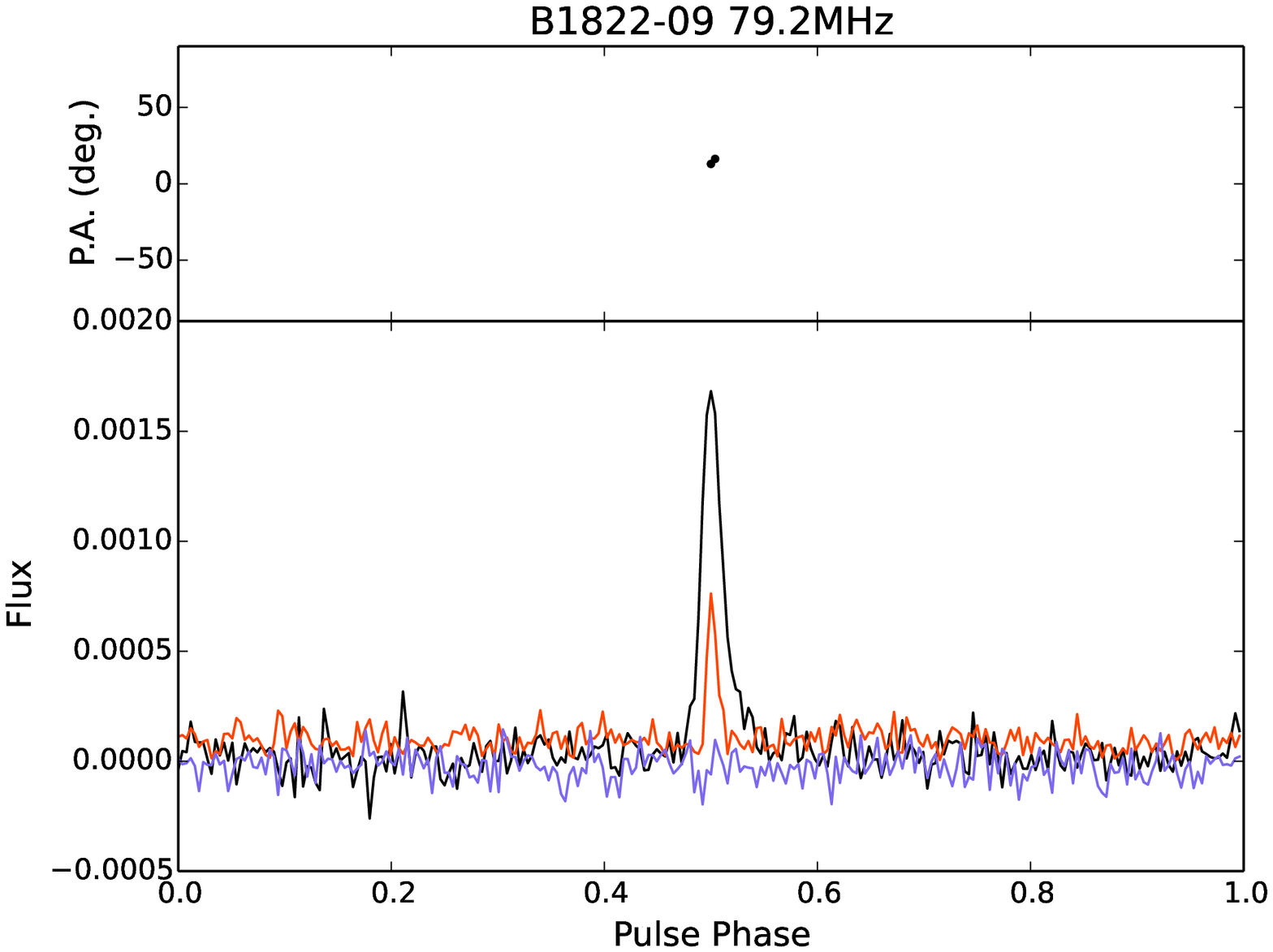}
	\caption{Profiles for B1822$-$09 at 79 MHz with colors as described in Fig.~ \ref{B0329fig}\label{B1822fig}}
\end{figure}

\begin{figure}
	\includegraphics[width=\columnwidth]{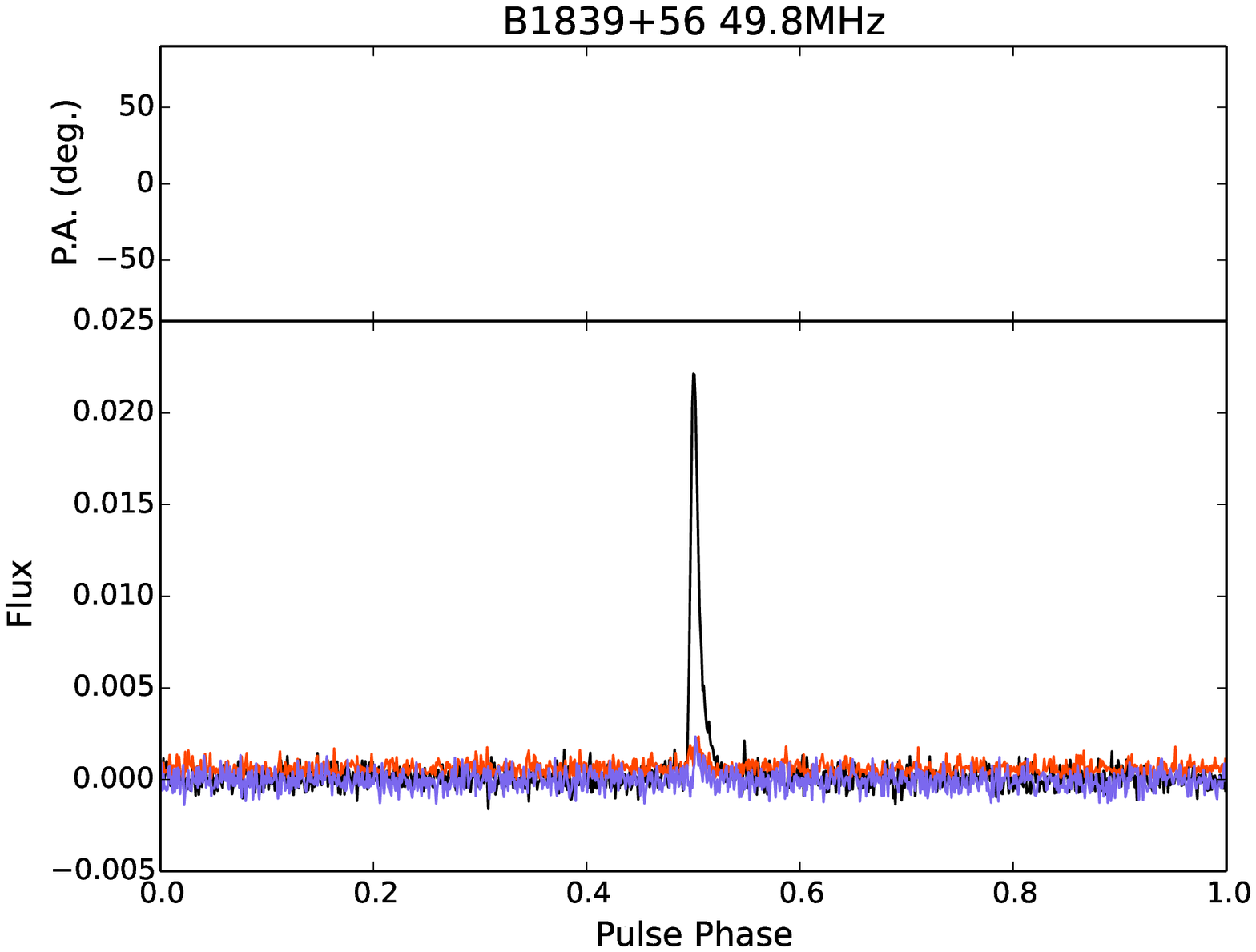}
    \includegraphics[width=\columnwidth]{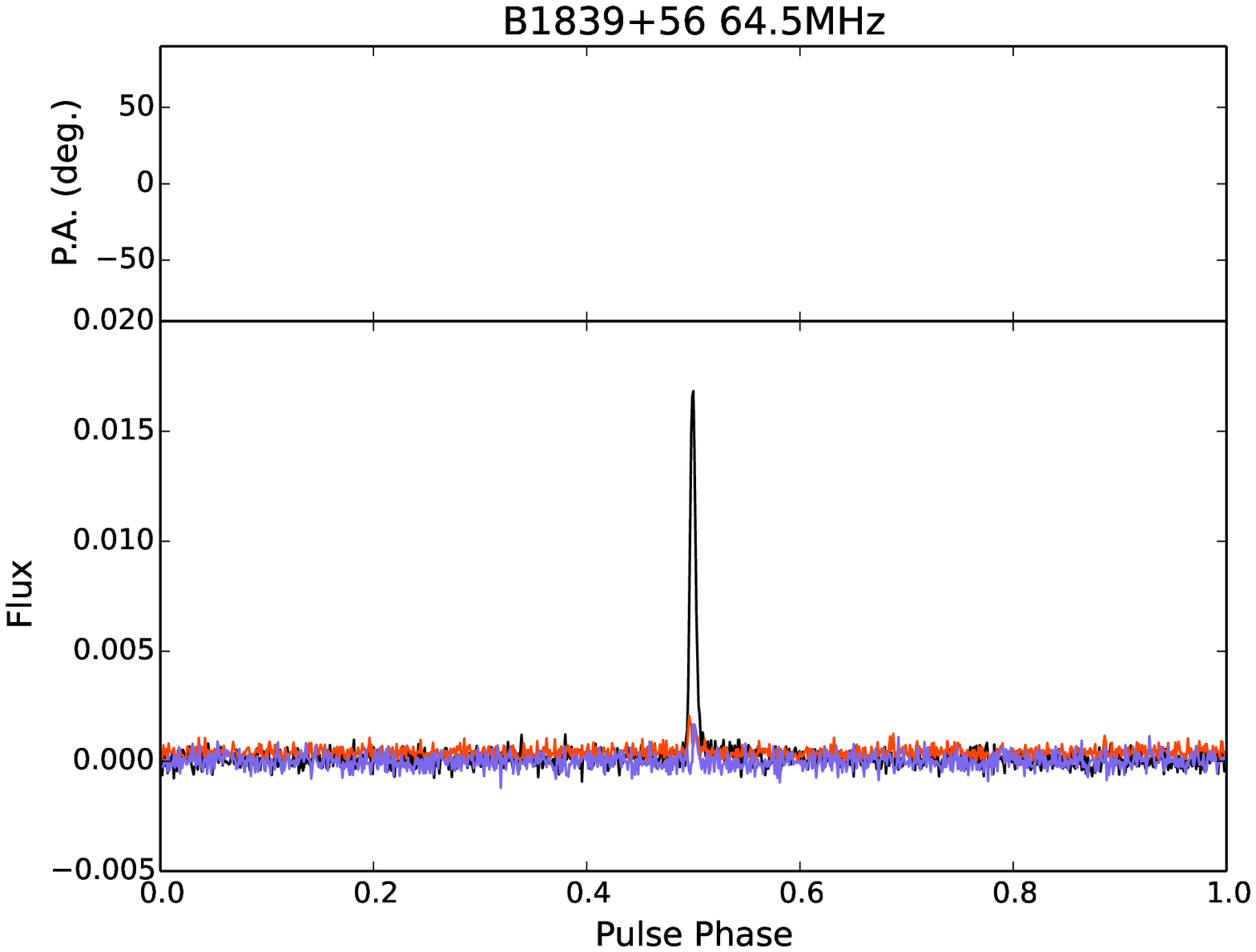}
    \includegraphics[width=\columnwidth]{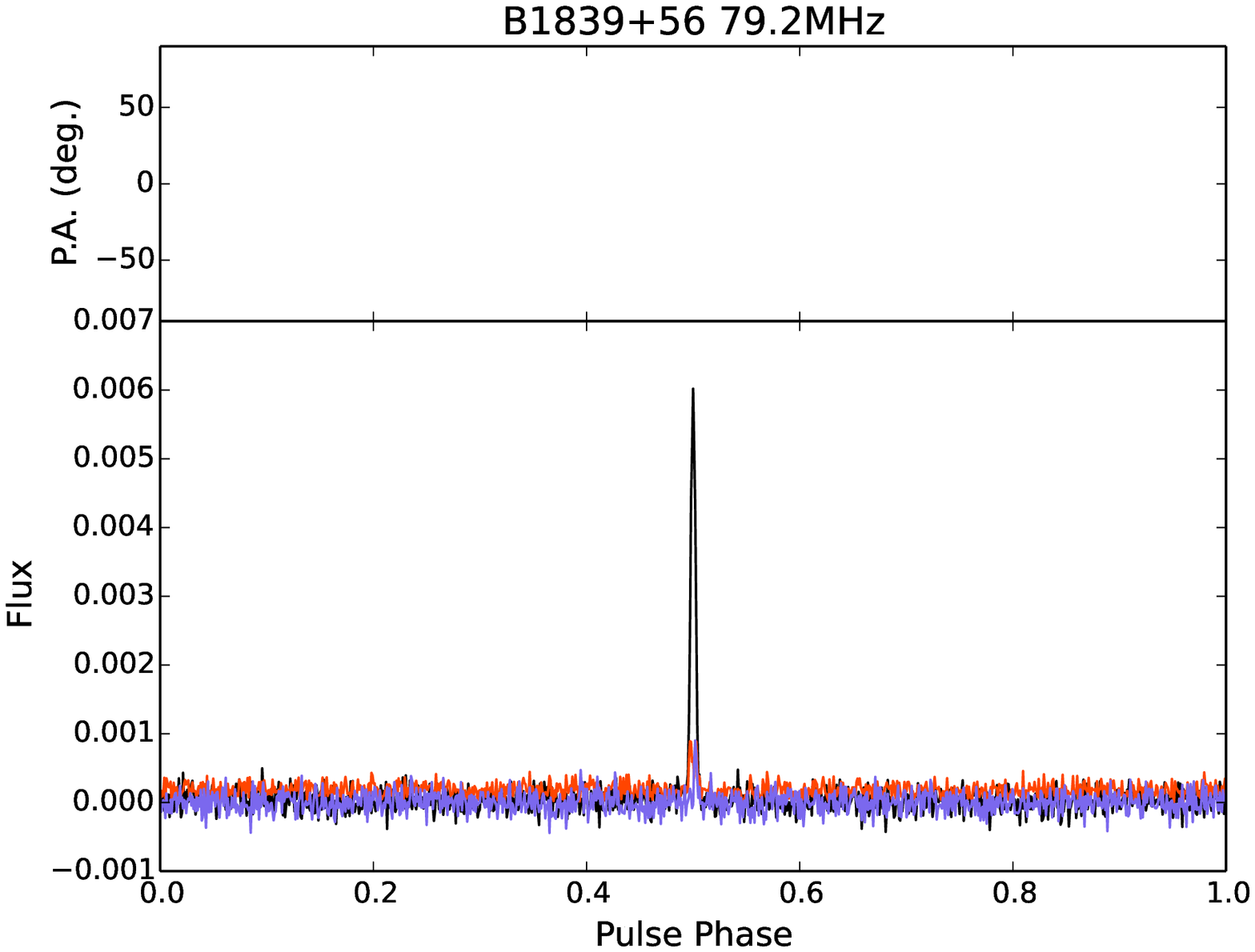}
	\caption{Profiles for B1839+56 at 49, 64 and 79 MHz with colors as described in Fig.~ \ref{B0329fig}\label{B1839fig}}
\end{figure}

\begin{figure}
	\includegraphics[width=\columnwidth]{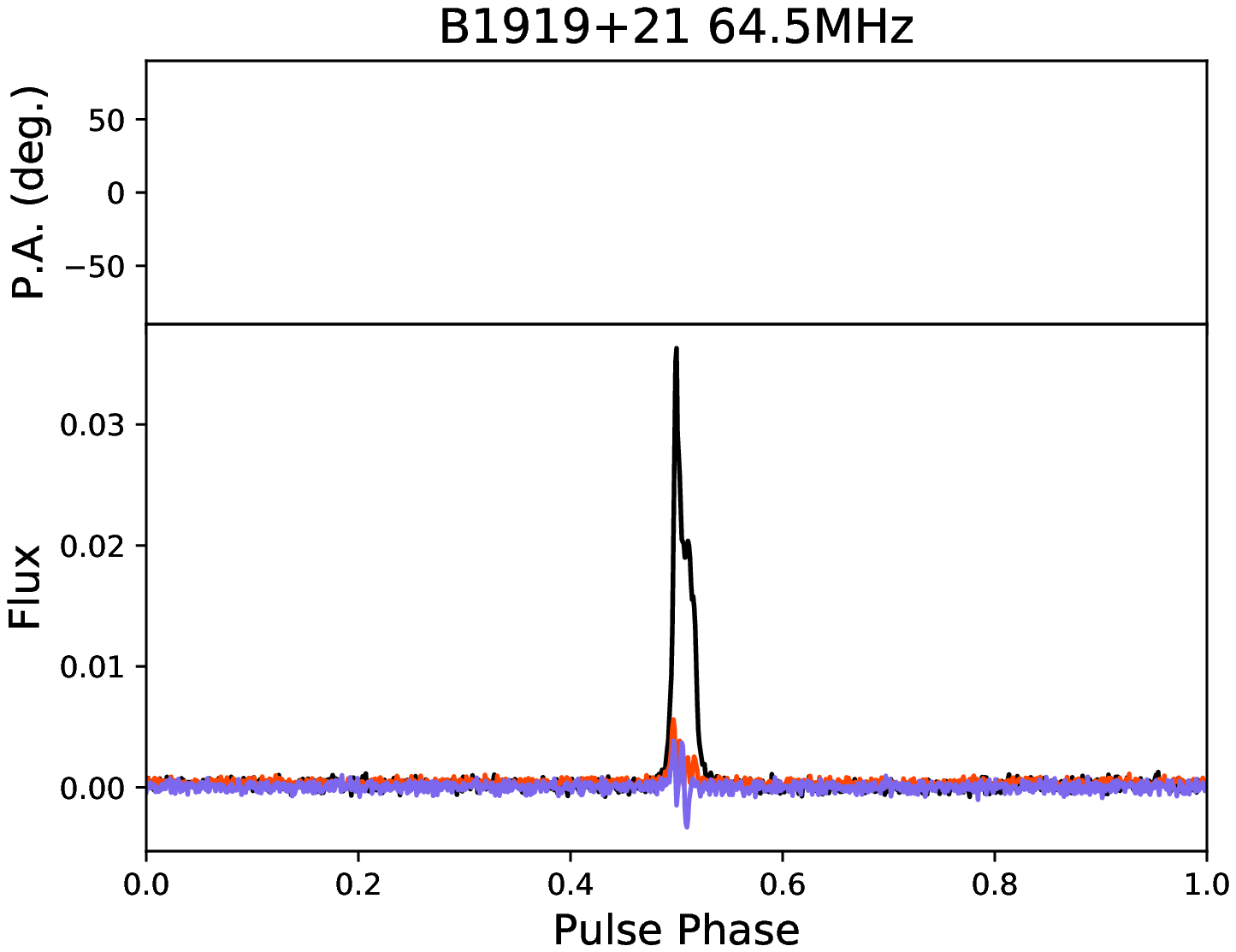}
    \includegraphics[width=\columnwidth]{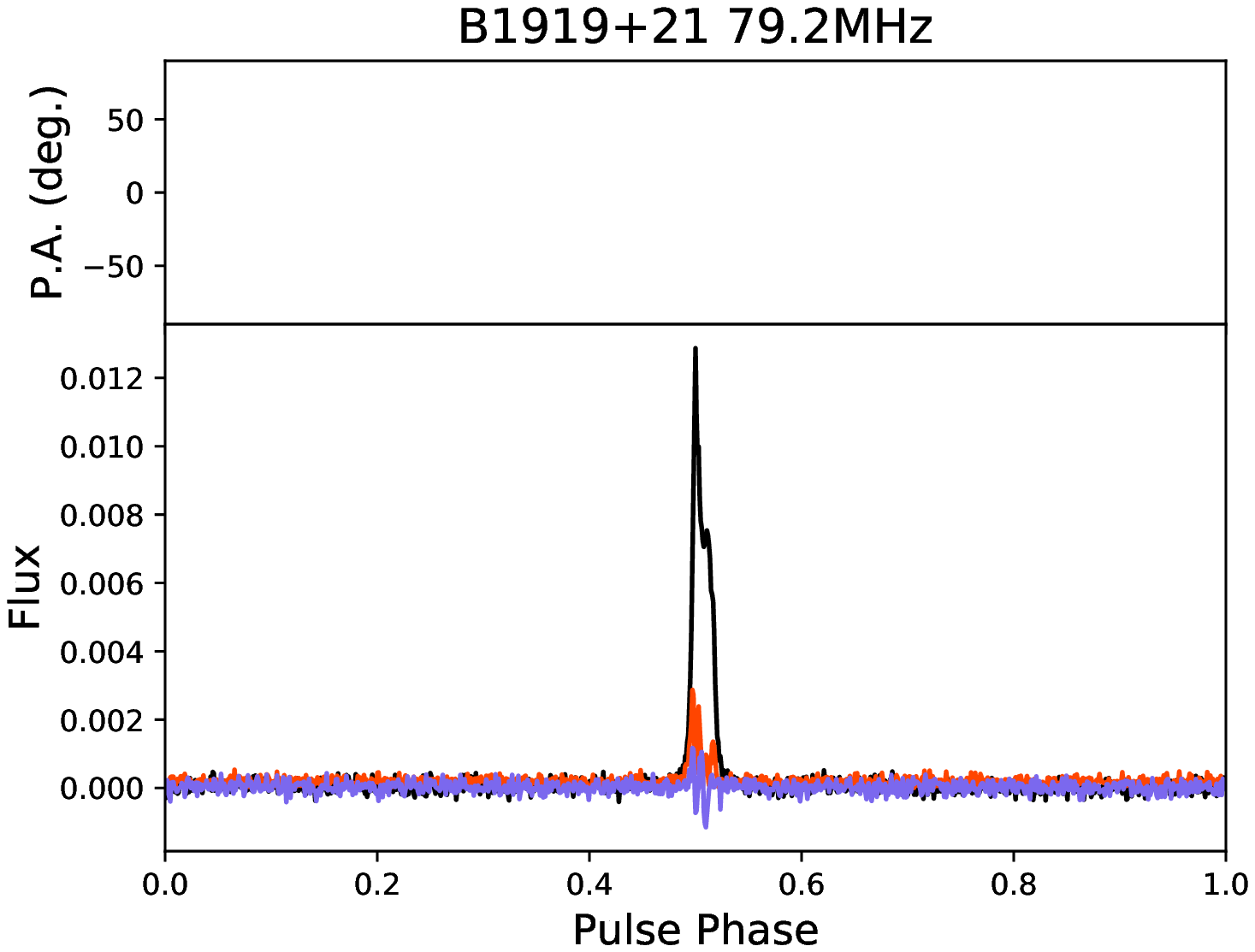}
	\caption{Profiles for B1919+21 at 64 and 79 MHz with colors as described in Fig.~ \ref{B0329fig}\label{B1919fig}}
\end{figure}

\begin{figure}
	\includegraphics[width=\columnwidth]{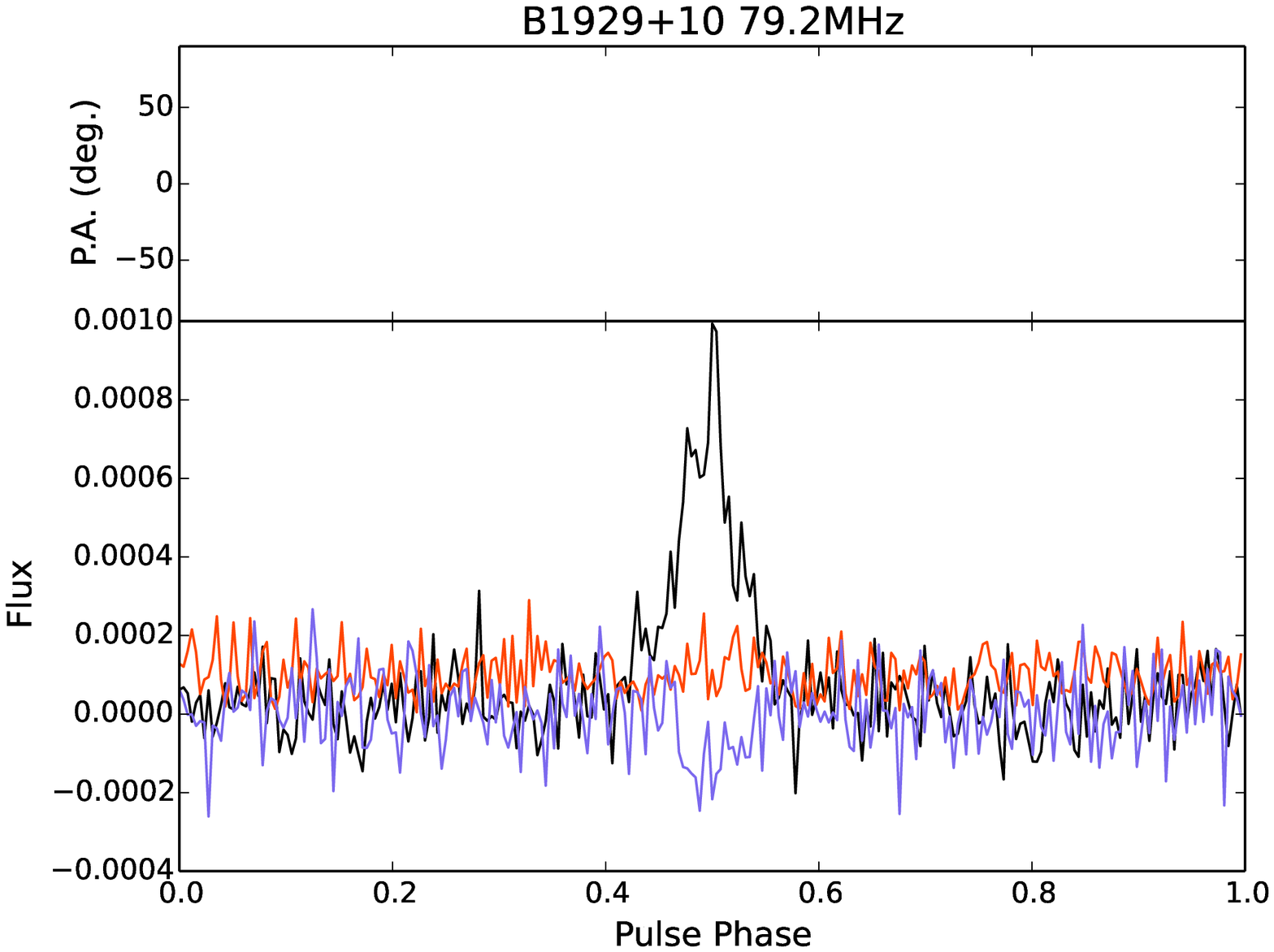}
	\caption{Profiles for B1929+10 at 79 MHz with colors as described in Fig.~ \ref{B0329fig}\label{B1929fig}}
\end{figure}

\begin{figure}
	\includegraphics[width=\columnwidth]{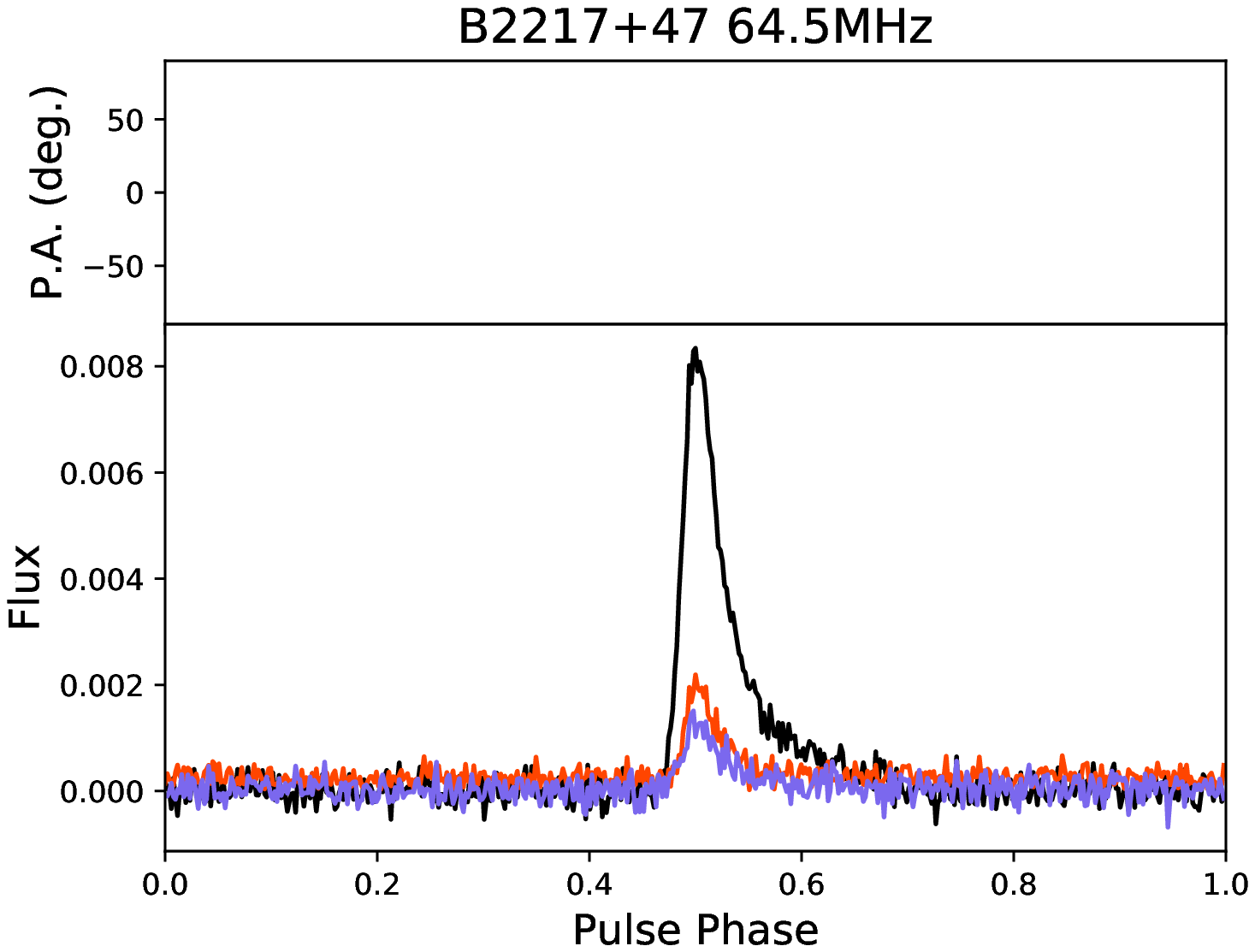}
    \includegraphics[width=\columnwidth]{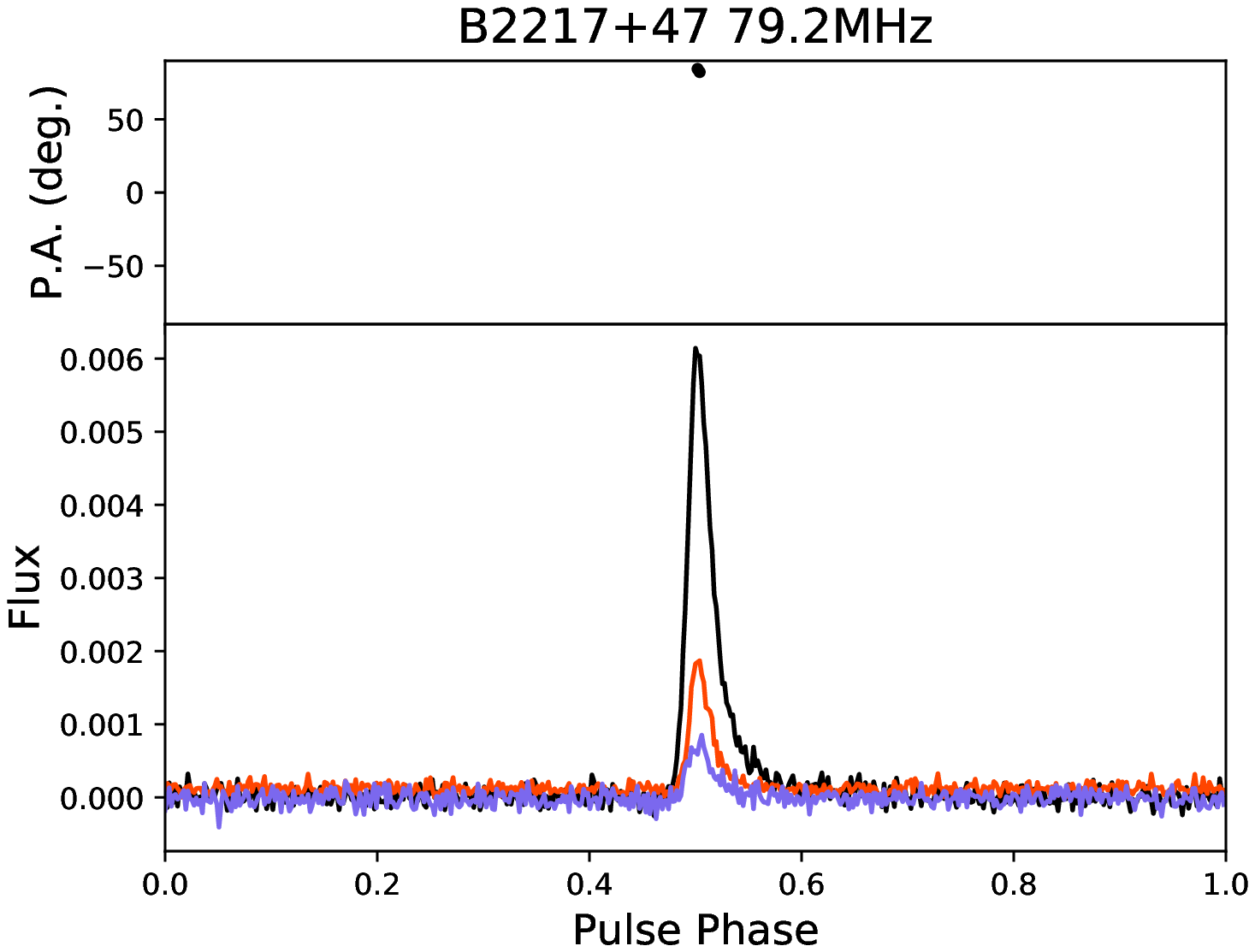}
	\caption{Profiles for B2217+47 at 64 and 79 MHz with colors as described in Fig.~ \ref{B0329fig}\label{B2217fig}}
\end{figure}

% Don't change these lines
\bsp	% typesetting comment
\label{lastpage}
\end{document}